\documentclass[11pt]{article}
\usepackage[utf8]{inputenc}
\usepackage{graphicx}
\usepackage{amsmath,amssymb,amsfonts}
\usepackage{color}
\usepackage{soul}
\usepackage{enumerate}
\usepackage{setspace}
\usepackage{caption}
\usepackage{geometry}
\usepackage{url}
\usepackage[section]{placeins}
\usepackage[font=small]{caption}
\makeatletter
\renewcommand{\@biblabel}[1]{\quad#1.}
\makeatother

\begin{document}

\begin{titlepage}
\title{The asexual genome of Drosophila}
\author{Stephan Schiffels${}^{1,*}$, Ville Mustonen${}^2$ and Michael Lässig${}^{3,*}$}
\date{\small{${}^1$ Department for Archaeogenetics, Max-Planck-Institut for the Science of Human History, Kahlaische Str. 10, 07749 Jena, Germany\\
${}^2$ Department of Biosciences, Department of Computer Science, Institute of Biotechnology, University of Helsinki, PO Box 65, 00014 Helsinki, Finland\\
${}^3$ Institut für Theoretische Physik, Universität zu Köln, Zülpicherstr. 77,50937, Köln, Germany\\
${}^*$ Corresponding Author
}}

\maketitle

\section*{Abstract}
The rate of recombination affects the mode of molecular evolution. In high-recombining sequence, the targets of selection are individual genetic loci; under low recombination, selection collectively acts on large, genetically linked genomic segments. Selection under linkage can induce clonal interference, a specific mode of evolution by competition of genetic clades within a population. This mode is well known in asexually evolving microbes, but has not been traced systematically in an obligate sexual organism. Here we show that the \emph{Drosophila} genome is partitioned into two modes of evolution: a \emph{local interference} regime with limited effects of genetic linkage, and an \emph{interference condensate} with clonal competition. We map these modes by differences in mutation frequency spectra, and we show that the transition between them occurs at a threshold recombination rate that is predictable from genomic summary statistics. We find the interference condensate in segments of low-recombining sequence that are located primarily in chromosomal regions flanking the centromeres and cover about 20\% of the \emph{Drosophila} genome. Condensate regions have characteristics of asexual evolution that impact gene function: the efficacy of selection and the speed of evolution are lower and the genetic load is higher than in regions of local interference. Our results suggest that multicellular eukaryotes can harbor heterogeneous modes and tempi of evolution within one genome. We argue that this variation generates selection on genome architecture.

\section*{Author Summary}
The \emph{Drosophila} genome is an ideal system to study how the rate of recombination affects molecular evolution. It harbors a wide range of local recombination rates, and its high-recombining parts show broad signatures of adaptive evolution. The low-recombining parts, however, have remained dark genomic matter that has been omitted from most studies on the inference of selection. Here we show that these genomic regions evolve in a different way, which involves clonal competition and is akin to the evolution of asexual systems. This regime shows a lower efficacy of selection, a lower speed of evolution, and a higher genetic load than high-recombining regions. We argue these evolutionary differences have functional consequences: protein stability and protein expression are gene traits likely to be partially compromised by low recombination rates.

\end{titlepage}


\section*{Introduction}
Genetic linkage affects molecular evolution by coupling the selective effects of mutations at different loci. This coupling, which is often called interference selection, generates two basic evolutionary processes: a strongly beneficial mutation can drive linked neutral and deleterious mutations to high frequency; conversely, a strongly deleterious mutation can impede the establishment of a linked neutral or beneficial mutation. The first process is an instance of hitchhiking or genetic draft \cite{Smith1974-pf,Wiehe1993-dc,Braverman1995-ef,Barton2000-nj,Fay2000-rh,Gillespie2000-fa,Gillespie2001-sj,Kim2000-je,Innan2003-rv,Kim2006-ni,Neher2011-dl,Schiffels2011-wn}, the second is known as background selection \cite{Charlesworth1995-dy,Charlesworth1993-nk,Charlesworth1994-lm,Charlesworth1996-xh,Hudson1995-kx,Nicolaisen2012-eb,Desai2013-nl,Good2014-bq,Campos2017-nt}. In both cases, interference selection has the same consequence: by spreading the effect of selected mutations onto neighbouring genomic sites, it reduces speed and degree of adaptation.

In an evolving population, interference links between genomic loci are established by new mutations, reinforced by selection on these mutants, and reduced by recombination. Hence, strength and genomic range of interference are set by all three of these evolutionary forces. In sexually reproducing organisms with a sufficiently high recombination rate, interference has limited effects because it remains local; that is, it acts only on mutations at proximal genomic positions but is randomized by recombination at larger distances. Without recombination, however, interference becomes global: it couples the evolution of mutations across an entire chromosome. Under strong selection pressures, chromosome-wide genetic linkage generates a specific mode of evolution in which populations harbour competing clades of closely related individuals, each clade containing a distinct set of beneficial and deleterious mutations. This evolutionary mode, which is commonly called clonal interference \cite{Gerrish1998-rx}, is well known in asexually evolving microbial \cite{Woods2011-mh} and viral populations \cite{Miralles1999-jo,Strelkowa2012-ts}. Theoretical models suggest that a similar mode of evolution, in which selection acts on extended segments of genetically linked sequence, arises in sexual populations at low recombination rates \cite{Weissman2012-of,Neher2013-qj}. However, finding evidence for this mode of evolution in the eukaryotic genome of an obligate sexual organism has remained elusive so far.

\emph{Drosophila} is an ideal system to map differential effects of interference in one genome. Fruit flies show overall high rates of adaptive genetic mutations \cite{Andolfatto2005-ij,Macpherson2007-yz,Mustonen2007-kq}, at least in high-recombining parts of the genome (low-recombining parts have been excluded from most previous studies). At the same time, local recombination rates vary by orders of magnitude within the \emph{Drosophila} genome. In particular, extended segments of low-recombining sequence are located in genomic regions flanking the centromeres and, to a lesser extent, the telomeres \cite{Comeron2012-jh}. In this paper, we present a systematic analysis of linkage effects across the \emph{Drosophila} genome. For two populations of \emph{D. melanogaster}, we map the frequency statistics of mutations and the divergence from the neighboring species \emph{D. simulans} in their dependence on the local recombination rate. Consistently in both populations, we find two clearly distinct regimes: a \emph{local interference} regime in high-recombining regions and an \emph{interference condensate} regime in extended low-recombing regions, which cover about 20\% of the autosomes. We delineate these two regimes by the statistics of synonymous mutations. In the local interference regime, the amount of synonymous mutations decreases with recombination rate and the frequency spectrum follows an almost perfect inverse-frequency power law, as predicted by classic theory \cite{Kimura1955-qy}. This indicates that the establishment of synonymous mutations is constrained by background selection \cite{Charlesworth1993-nk,Charlesworth1994-lm,Charlesworth1995-dy,Charlesworth1996-xh,Nicolaisen2012-eb}, but established mutations evolve predominantly under genetic drift. In contrast, the frequency spectrum of the interference condensate regime shows a specific depletion of intermediate- and high-frequency variants, which is consistent with genetic draft over extended genomic distances~\cite{Gillespie2000-fa,Gillespie2001-sj}.

To corroborate this scenario, we develop a scaling theory of evolution under positive and negative selection and limited recombination, building on recent models of asexual and sexual evolution \cite{Gerrish1998-rx,Neher2011-dl,Neher2013-qj,Desai2007-yb,Park2007-rf,Park2010-sz,Good2014-bq,Rouzine2008-jr}. This theory provides a unified framework for background selection and genetic draft. It describes how amount and frequency-dependence of mutations depend on the recombination rate, and it predicts the transition point from local interference to the interference condensate. Over a wide range of evolutionary parameters, the transition occurs at a threshold recombination rate that is close to the sum of the rates of deleterious mutations and of beneficial substitutions per unit sequence; these rates can be inferred from genomic summary statistics. In the \emph{Drosophila} genome, the predicted threshold recombination rate is numerically close to the point mutation rate, in perfect agreement with the transition point observed in mutation frequency spectra.

Our scaling theory also provides the tools to infer key evolutionary features of the condensate regime from genomic data. We use this inference to quantify similarities of the \emph{Drosophila} interference condensate to asexual evolution, and to understand the likely biological impact of the condensate mode of evolution. Specifically, we show that genes in condensate regions are less evolvable in response to positive selection and have a higher genetic load than genes in the local interference regime. We discuss how these evolutionary differences can impact gene functions in the condensate and generate selective pressure on genome architecture.

\section*{Results}
\subsection*{Evolutionary modes in recombining genomes}

Why there are two distinct evolutionary regimes in recombining genomes can be understood from a remarkably simple scaling theory. We consider genome evolution under deleterious mutations with rate $u_d$, beneficial mutations generating substitutions with rate $v_b$, and recombination with rate $\rho$; all of these rates are measured per base pair unit of haploid sequence and per generation. In terms of these rates, we estimate the probability that a given selected mutation evolves autonomously ---i.e., free of background selection and genetic draft---, generalizing a previous argument by Weissman and Barton for evolution solely under beneficial mutations~\cite{Weissman2012-of}. We first compute the average amount of interference generated by the substitution of a beneficial mutation with selection coefficient $s$ at a given focal site in the genome. This mutation takes an average time of order $\tau_s \sim 1/s$ from establishment to high frequency and generates a linkage correlation interval of size $\xi_s = 1/(\rho \tau_s) \sim s/\rho$ around the focal site. Other mutations at a distance $r \lesssim \xi_s$ are likely to retain their genetic linkage to one of the alleles at the focal site and are subject to strong interference; more distant mutations are likely to randomize their genetic linkage to the alleles at focal site by recombination within the time span $\tau_s$. Hence, each beneficial substitution generates an \emph{interference domain} with an area $\tau_s \times \xi_s \sim 1/\rho$ around its focal point in genomic space-time~\cite{Weissman2012-of}. By exactly the same argument, each deleterious mutation creates background selection in an interference domain with area $\tau_s \times \xi_s = 1/\rho$ around its focal point. In this case, $\tau_s \sim 1/s$ is the expected time between origination and loss of the deleterious allele. While genetic draft acts on all other mutations in the genomic neighborhood of the driver mutation, background selection strongly affects only mutations on the genetic background of the deleterious allele, but this difference does not affect the scaling of interference domains. To estimate the joint effects of beneficial and deleterious mutations, we combine both kinds of interference interactions into a single \emph{interference density} parameter
\begin{equation}
\omega = \frac{u_d + v_b}{\rho}.
\label{omega}
\end{equation}
This parameter delineates two universal evolutionary modes of recombining genomes:

\begin{figure}
  \includegraphics[width=\textwidth]{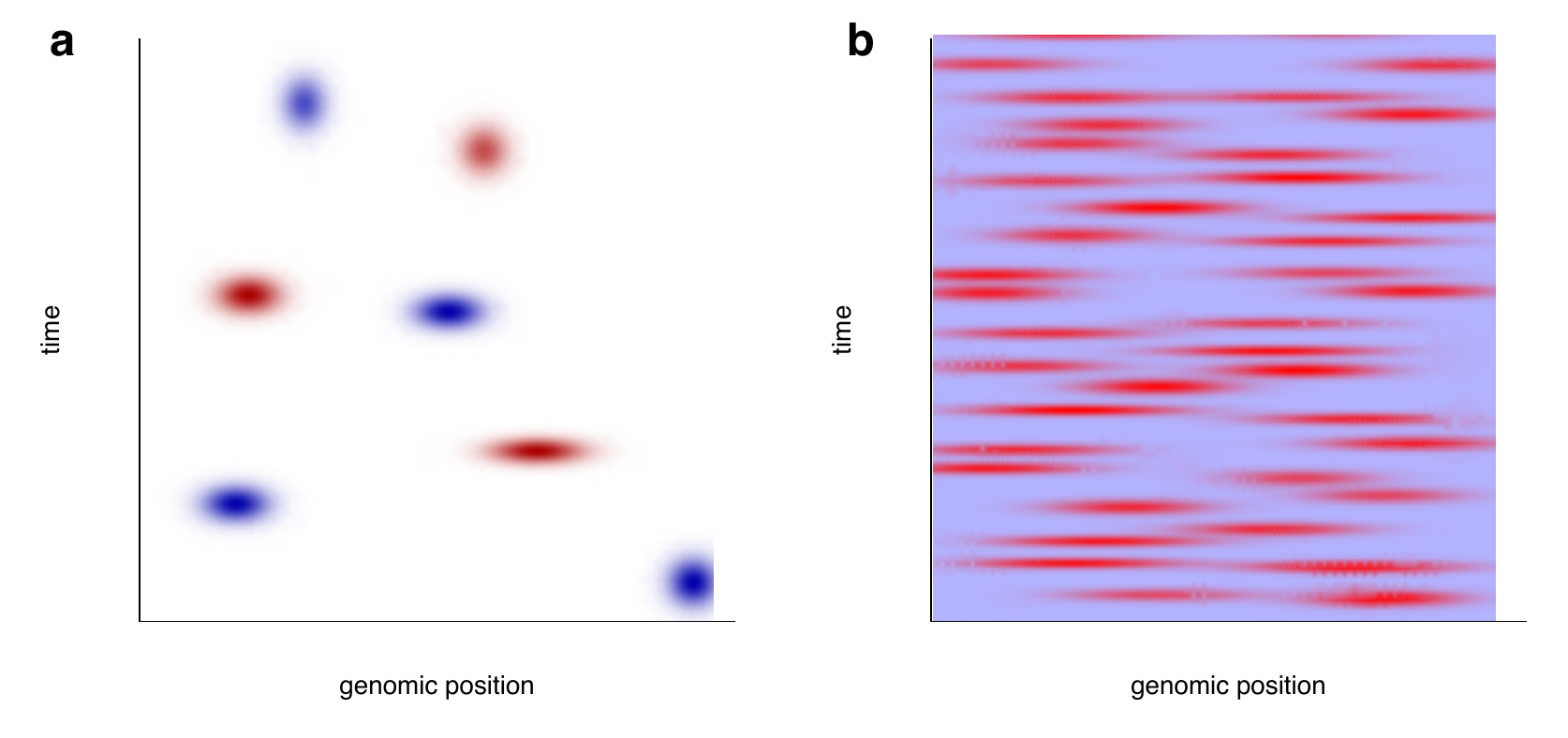}
  \caption{\textbf{Evolutionary modes in recombining genomes (schematic).} The interference density $\omega$, defined in equation~(\ref{omega}), delineates two regimes that differ in strength and genomic range of interference selection. (a) Local interference ($\omega \lesssim 1$). This regime has a dilute pattern of interference domains generated by deleterious mutations (blue) and beneficial substitutions (red). These domains have space-time densities $u_b$ and $v_b$, respectively; each domain covers a space-time area $1/\rho$. Interference selection occurs only within interference domains; mutations under selection evolve in a largely independent way. (b) Interference condensate ($\omega \gtrsim 1$). In this regime, interference domains generate a densely packed pattern. Interference domains have characteristic sizes $\xi = \tilde \sigma/ \rho$ and $\tau = 1/\tilde \sigma$ in space and time, as detailed in equation~(\ref{tauxi1}). Mutations are subject to strong interference selection, which curbs the speed of evolution.}
  \label{fig_schematic}
\end{figure}

The \emph{local interference} mode ($\omega \ll 1$) has a dilute pattern of interference domains: the domains of beneficial substitutions and of deleterious mutations are randomly distributed with probabilities $u_b$ and $v_b$ per unit sequence and per unit time (Fig.~\ref{fig_schematic}a).
The space-time shape of a domain, which is given by the scales $\xi_s$ and $\tau_s$, depends on the selection coefficient of its focal mutation, but its area $1/\rho$ is universal. Because the interference domains are, on average, well-separated in the local intereference mode, different mutations under selection are statistically independent. Hence, the rate of beneficial substitutions with selection coefficient $s$ is related to the underlying mutation rate $u_b$ and the haploid effective population size $N$ by Haldane's classic formula for individual sites, $v_b = 2Ns u_b$ \cite{Haldane1927-ep}. A given (beneficial or deleterious) mutation evolves autonomously if its own interference domain has negligible overlap with any of the other interference domains, which happens with probability $p_0 = {\rm e}^{-\omega}$. Two target mutations on different genetic backgrounds see the same red domains but different blue domains; however, the no-interference probability $p_0$ is independent of background.

The \emph{interference condensate} mode ($\omega \gg 1$) has densely packed and overlapping interference domains. This indicates strong interference over extended genomic segments: genomic space-time is jammed by mutations under selection (Fig.~\ref{fig_schematic}b). By the definition of $\omega$, the condensate is a broad evolutionary regime: it is generated by a sufficiently large supply of substantially beneficial or deleterious mutations, or a combination of both, but it does not depend on details of their effect distribution. Remarkably, the condensate domains have not only a universal area but also a typical shape that is given by universal scales $\tau$ and $\xi = 1/(\rho \tau)$. Below, we will infer these scales from genomic data in \emph{Drosophila}.

\begin{figure}
  \includegraphics[width=\textwidth]{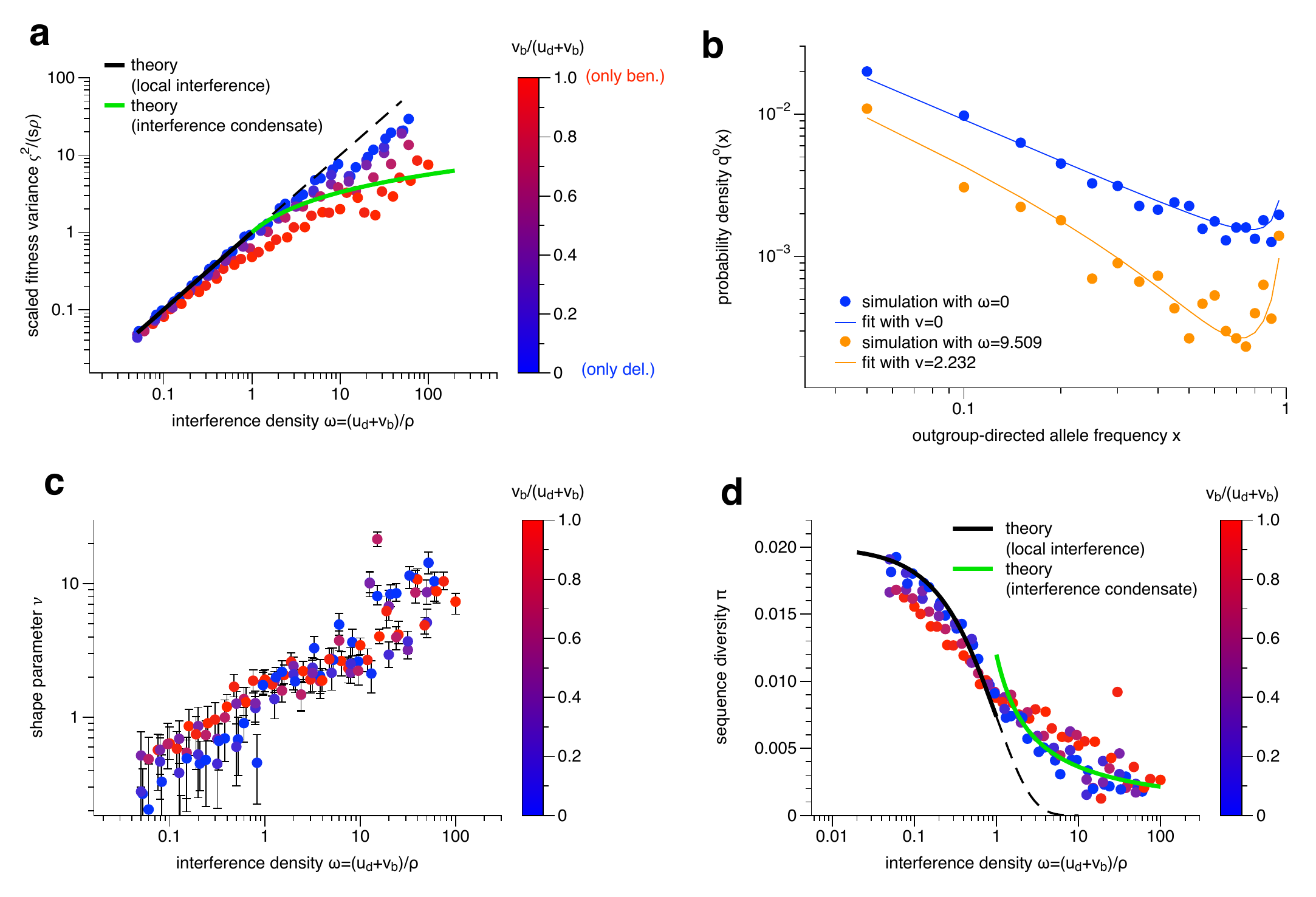}
  \caption{\textbf{Genomic markers of interference.} Simulations of evolution with beneficial mutations (rate $u_b$ per bp, selection coefficient $s$), deleterious mutations (rate $u_d$ per bp, selection coefficient $-s$), and recombination (rate $\rho$ per bp) in haploid populations of size $N$.
  (a) Scaled fitness variance per site, $\varsigma^2 / (s \rho)$, plotted against the interference density $\omega = (u_d + v_b)/\rho$ with $v_b = 2Ns u_b$. Simulation data for different values of $u_d$ and $u_b$ (marked by the color of points) are shown together with theoretical predictions for the local interference regime ($\omega \lesssim 1$, black line) and the interference condensate ($\omega \gtrsim 1$, green line). See equations (\ref{varsigma2}) and (\ref{eq:sigma_condensate}).
  (b) Frequency spectrum of neutral mutations. Simulation results for parameters in the local interference regime (blue dots) and in the interference condensate (orange dots) are shown together with fits to the spectral functions $Q_0(x; \nu \! = \! 0)$ (blue line) and  $Q_0(x; \nu \! = \! 2.7)$ (orange line). The shape distortion at intermediate frequencies is a universal characteristic of the condensate regime. The secondary branch at high frequencies is a feature of outgroup-directed spectra, which are appropriate for our data analysis; see text below and equations~(\ref{Q0}) and~(\ref{qso}).
  (c) Shape parameter $\nu$, plotted against interference density $\omega$. Site frequency spectra from simulations are fitted to the spectral function $Q_0(x; \nu)$.
  (d) Neutral sequence diversity $\pi$ plotted against interference density $\omega$. Simulation data is shown together with theoretical predictions for the local interference regime ($\omega \lesssim 1$, black line) and the interference condensate ($\omega \gtrsim 1$, green line). See equations~(\ref{pi}) and (\ref{eq:pi_condensate}).
  Simulation parameters: $N=2000$,  $\rho = 10^{-7}-10^{-4}$, $u_d = 0-3.0\times 10^{-6}$, $u_b = 0-2.5\times 10^{-7}$.}
  \label{fig_sim}
\end{figure}

A key evolutionary quantity to map these evolutionary modes is the average fitness variance in a population, $\Sigma$, which measures the efficacy of selection: by Fisher's fundamental theorem, $\Sigma$ equals the rate of fitness increase by frequency gains of fitter genetic variants in the population. As detailed in Methods, our scaling theory captures the dependence of $\Sigma$ on the evolutionary parameters in both interference modes. In the local interference regime, $\Sigma$ depends linearly on the rates $u_d$ and $u_b$, which reflects the statistical independence of mutation events. In the interference condensate, $\Sigma$ is curbed to a sublinear function of $u_d$ and $u_b$, which signals a reduced efficacy of selection caused by the jamming of genomic space-time (Fig.~\ref{fig_schematic}b). To test these results of our scaling theory, we performed simulations of evolving populations (see Methods for simulation details). In Fig.~\ref{fig_sim}a, the average fitness variance per unit sequence, $\varsigma^2$, is plotted against $\omega$ (with a suitable rescaling factor, as explained in Methods). In the regime $\omega \ll 1$, the rescaled fitness variance data obtained for a wide range of parameters $u_d$, $u_b$, $\rho$ collapse onto a uni-valued, linear function of $\omega$. In the interference condensate ($\omega \gtrsim \omega^*$), the fitness variance is seen to be curbed, and the rescaled data for different evolutionary parameters show some spread. Here we evaluate $\omega$ in terms of the mutation rates and the average selection coefficient of beneficial mutations, $\omega  = (u_d + 2N \bar s u_b)/\rho$, which maps the same regimes as equation~(\ref{omega}) because $v_b \simeq 2N \bar s u_b$ for $\omega \lesssim 1$.

The reduction in the efficacy of selection and, in particular, the diminishing return of beneficial mutations in the interference condensate are hallmarks of asexual evolution in large populations, which are known from experiments with microbial populations \cite{Wiser2013-wl,Woods2011-mh,Lang2013-hu,Perfeito2007-lj} and from theoretical models \cite{Gerrish1998-rx,Rouzine2008-jr,Schiffels2011-wn,Desai2007-yb}. They signal the competition between genetic clades in an evolving population, which prevents some beneficial mutations to reach substantial frequencies because they are outrun by other clades. At the same time, deleterious mutations can reach fixation if they are part of a successful clade; this effect is often referred to as Muller’s ratchet \cite{Muller1932-lk,Muller1964-vh,Jain2008-js,Mustonen2009-st,Goyal2012-ro,Neher2013-xf}. We conclude that the condensate regime shares important characteristics with asexual processes, in accordance with previous results in refs.~\cite{Weissman2012-of, Neher2013-qj,Neher2011-dl}. Our scaling theory expresses this link mathematically: the modes of evolution in recombining systems can be mapped onto corresponding modes of asexual evolution in a specific low-recombination limit (Methods).

In a minimal scaling theory that is based on the interference density, the local interference regime and the interference condensate are separated by a smooth transition at a characteristic value $\omega^*$ of order 1. The transition occurs when the interference probability $e^{-\omega}$ becomes of order 1; the transition point marks the onset of nonlinearity in the fitness variance. For a broad range of evolutionary parameters, which includes realistic assumptions for the \emph{Drosophila} genome, this behavior is confirmed by our simulations (Fig.~\ref{fig_sim}a) and is consistent with analytical results for specific cases~\cite{Weissman2012-of,Neher2013-qj}, including the low-recombination limit~\cite{Desai2007-yb,Good2014-bq,Schiffels2011-wn}. In Methods, we detail this minimal scaling theory and discuss extensions that cover parameter regimes with systematic shifts of the transition point $\omega^*$ above or below 1 (Supplementary Figure S1).

\subsection*{Genomic signature of interference selection}

The fitness variance is a key summary statistics to map evolutionary modes under interference selection but it depends on the {\em a priori} unknown rates $u_d$ and $v_b$. The distribution of mutation frequencies $x$ at neutrally evolving sites, the so-called neutral site frequency spectrum $q (x)$, provides an alternative test that can be directly evaluated from population-scale sequencing data. At high recombination rates, the neutral spectrum is dominated by genetic drift and has the universal Kimura form $q (x) \sim 1/x$  \cite{Kimura1955-qy} (blue line in Fig.~\ref{fig_sim}b). At low recombination, the spectrum shows a characteristic depletion of intermediate- and high-frequency mutation counts (red line in Fig.~\ref{fig_sim}b). This shape distortion is a result of genetic draft, which generates faster frequency changes and, hence, fewer variants in this frequency range than genetic drift. This distortion turns out to be a robust feature that can be read off from genomic data even if the exact form of the spectrum is hidden by noise and confounding factors.

We can compare the site frequency spectrum inferred from genomic data with spectra derived from specific evolutionary models.
All analytically solvable models make strong simplifying assumptions on the evolutionary process, specifically on rate and effect distributions of mutations generating interference selection. The exact form of the spectral function depends on these model details, but broad shape features are a universal markers of interference. An important class of models are so-called travelling fitness waves, which describe the asymptotic regime of linked genetic variation generated by multiple coexisting mutations with individually small selection coefficients \cite{Desai2007-yb,Park2007-rf,Rouzine2008-jr,Hallatschek2011-uu}. Fitness waves generate a steady turnover of sequence variation at a characteristic rate $\sigma$. In the asymptotic wave regime, genetically linked neutral sites have a spectrum depleted at intermediate frequencies, as given by an inverse-square power law $q (x) \sim 1/x^2$ for $x<1/2$ and a minimum near $x = 1/2$ \cite{Neher2013-qj}. These models underscore an important general point: genetic draft --- i.e., mutation frequency trajectories shaped by the substitution dynamics of a beneficial allele at a genetically linked locus --- and the associated shape distortion of site frequency spectra is universally generated by a sufficient supply of deleterious or by beneficial mutations~\cite{Goyal2012-ro, Neher2013-xf}. In the following, we use a specific model with spectral shapes depleted at intermediate frequencies that are tunable to the \emph{Drosophila} data reported below. The model contains a focal site that evolves by mutations, selection, and genetic drift; the site is also subject to background selection and genetic draft with rate $\sigma$. Draft is generated by linked strongly beneficial alleles, each of which occurs on a random genetic background and leads to instantaneous fixation or loss of mutations at the focal site. The resulting spectral function of neutral sites takes a simple approximate form, $Q_0 (x; \nu) = {\rm e}^{-\nu x} /x$, with a shape parameter $\nu$ that is proportional to the draft rate $\sigma$ (dashed lines in Fig.~\ref{fig_sim}b; details on $Q_0 (x; \nu)$ are given in Methods). In the following, this model will serve to parametrize the universal shape distortion of empirical spectra, without pretence to resolve details of the underlying selective forces.

In Fig.~\ref{fig_sim}cd, we show that mutation frequency data produce two consistent markers of interference. First, we fit neutral site frequency spectra obtained from numerical simulations to the form $Q_0 (x; \nu)$ and plot the inferred shape parameter $\nu$ against the interference density $\omega$ of the underlying evolutionary process. Over a wide range of rates $u_s, u_b$, and $\rho$, we find $\nu \approx 0$ (i.e., neutral spectra of the form $q (x) \sim 1/x$) in the local interference regime ($\omega \lesssim \omega^*$) and $\nu > 1$ (i.e., neutral spectra with depletion of intermediate and high frequencies) in the condensate regime ($\omega \gtrsim \omega^*$). Below, we will link the shape parameter $\nu$ to specific evolutionary characteristics of the condensate regime. Second, we record the sequence diversity at synonymous sites as a function of the interference density $\omega$ (Fig.~\ref{fig_sim}d). This quantity shows a strong dependence on $\rho$ in the local interference regime ($\omega \lesssim \omega^*$) and a weaker dependence in the interference condensate ($\omega \gtrsim \omega^*$), a pattern that is predicted by our scaling theory and will be described in more detail below. Hence, both genetic draft on neutral mutations and the depletion of the diversity pattern set on at the transition point $\omega^*$ from local interference to the interference condensate (Fig.~\ref{fig_sim}cd), the same point that is marked by the onset of nonlinearity in the fitness variance (Fig.~\ref{fig_sim}a). The validity range of our inference method is detailed in Methods.

\subsection*{Interference selection in the \emph{Drosophila} genome}

To obtain a genome-wide map of interference in the \emph{Drosophila melanogaster} genome, we use sequence data from an American \cite{Mackay2012-yt} and an African population \cite{Pool2012-fy}. To equalize coverage, we take a random sample of 25 individuals in each population. At this sampling depth, site frequency spectra are quite insensitive to low-frequency variants (which would arise, for example, from a recent population expansion) but are perfectly suitable for studying intermediate-frequency variants, which are at the center of this study. Based on a published high-resolution recombination map for the Drosophila genome \cite{Comeron2012-jh}, we partition genomic sites in the autosomes (i.e., chromosomes 2L, 2R, 3L, 3R) according to the local recombination rate evaluated in windows of $10^5$ base pairs. This partitioning covers a range of rates between $10^{-10}$ and $10^{-7}$ with an average of $2.4 \times 10^{-8}$ per unit sequence and per generation (often reported in units of $10^{-8}$  per unit sequence and per generation, called centiMorgans per Megabase).

In each recombination rate bin and in different sequence categories, we record the outgroup-directed site frequency spectrum, $\hat q^o (x)$, which is defined as the number of sites per unit sequence at which a fraction $x = k/n$ of the sampled individuals have a mutant allele and a fraction $ 1 - x$ have the outgroup allele (with $k = 0,1, \dots, 25$ and $n = 25$; we disregard sites with more than two alleles). Following common practice, we determine the outgroup allele by alignment with the reference genome of the neighboring species {\em D. simulans}. These empirical spectra differ in two ways from the model spectra introduced above. First, the spectra $\hat q^o (x)$ are evaluated for discrete frequencies in a small sequence sample, which introduces sampling corrections compared to model distributions derived for larger populations (we use a hat to mark this difference; sampling corrections are detailed in Methods). Second, model spectra are directed from the ancestral allele at the origination time of the mutation. A substitution between the ingroup and the outgroup species reverses the role of the ancestral and allele mutant at part of the sequence sites. In a sequence class with a given density $d$ of substitutions, the ancestor-directed and the outgroup-directed spectra are related by a linear map, $q^o (x) = (1-d) \, q (x) + d \, q (1-x)$; the same map relates the sample spectra $\hat q^o (x)$ and $\hat q (x)$. Hence, outgroup-directed spectra have a primary branch with a maximum at low frequency and a secondary branch with a maximum at high frequency.

\begin{figure}
  \includegraphics[width=\textwidth]{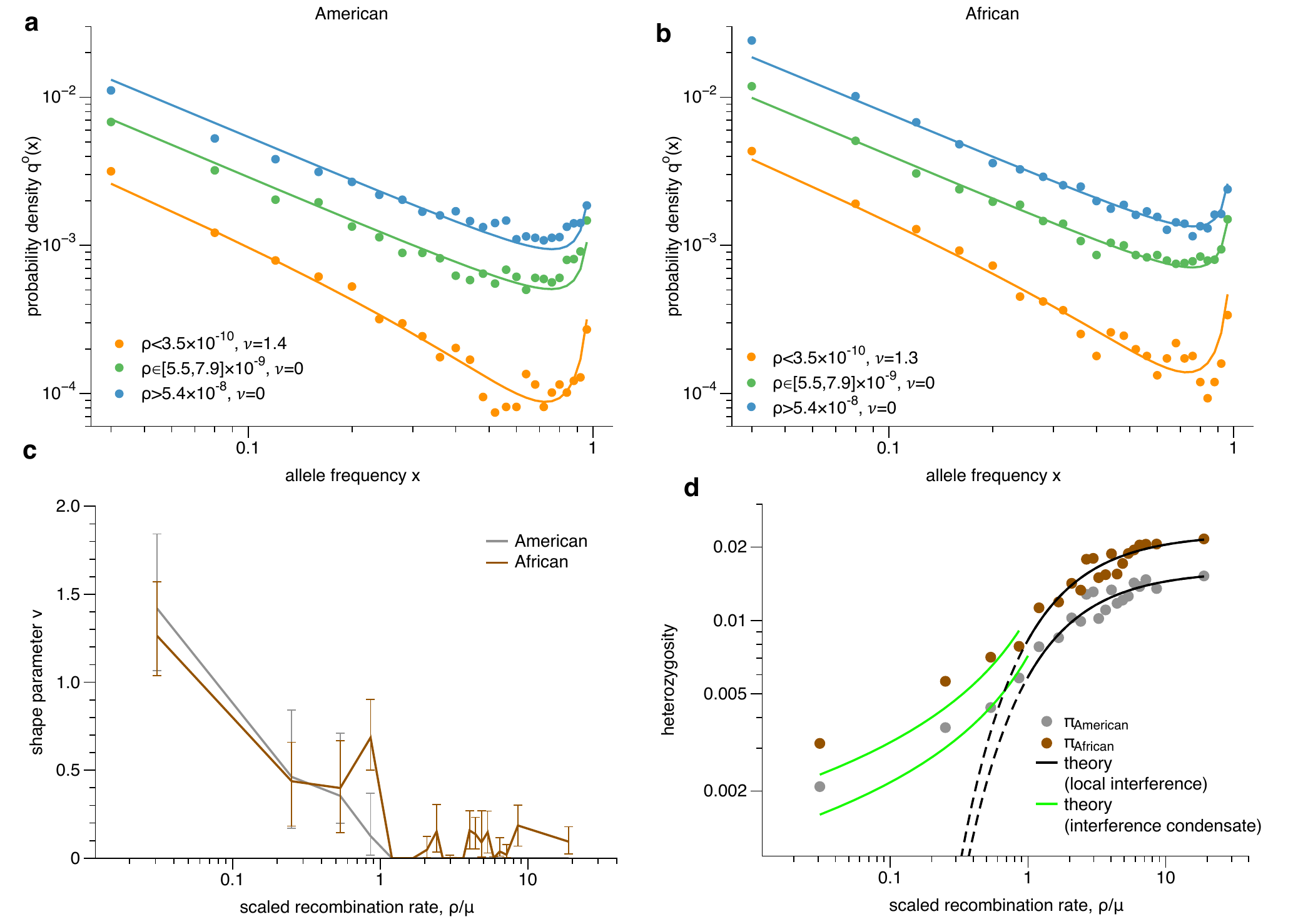}
  \caption{\textbf{Interference regimes in the \emph{Drosophila} genome.} (a,b) Outgroup-polarized frequency spectra of synonymous mutations. Sample spectra $\hat{q}_s^o(x)$ from two populations of \emph{D. melanogaster} at three values of the recombination rate $\rho$, together with model spectra $q_s^o(x)$ with maximum-likelihood shape parameter $\nu$ as indicated (solid lines). See equations~(\ref{qs}) and~(\ref{qso}). (c) Spectral shape of synonymous mutations. Inferred shape parameters $\nu$ from two populations are plotted against the scaled recombination rate $\rho/\mu$. (d) Sequence diversity at synonymous sites, $\pi_s$, plotted against the scaled recombination rate $\rho/\mu$ (dots). Data from two populations is shown together with theoretical predictions for the local interference regime ($\omega \lesssim 1$, black line) and the interference condensate ($\omega \gtrsim 1$, green line). See equations~(\ref{pi}) and (\ref{eq:pi_condensate}). The data of (c,d)  consistently maps the condensation transition to a threshold value $\rho^*/\mu \sim 1$, as predicted by scaling theory.}
  \label{fig_syn}
\end{figure}

In Fig.~\ref{fig_syn}ab, we plot the sample spectra of 4-fold synonymous sequence sites, $\hat q_s^o (x)$, for three representative bins of high, intermediate, and low recombination rates in the American and the African population. These spectra  have a striking common pattern. Across high and intermediate recombination rates, they follow almost perfectly the standard Kimura inverse-frequency form, $q (x) \sim (1-d)/x + d / (1-x)$, which appears as straight lines over most of the frequency range in the double-logarithmic plots of Fig.~\ref{fig_syn}ab. This form indicates that the dominant evolutionary force acting on synonymous genetic mutations at high and intermediate recombination rates is genetic drift. It shows that the average selection at synonymous sites is weak, making this class a good approximation of neutrally evolving sequence. It also excludes strong demographic effects affecting the spectral form at intermediate frequencies (notwithstanding small differences at low frequencies between the American and the African population, which reflect differences in their recent demography \cite{Stephan2007-zr,Duchen2013-cf}). Strong selective sweeps are known to deplete the density and to distort the spectrum of synonymous mutations in the local vicinity of the positively selected site \cite{Macpherson2007-yz,Sattath2011-nq}; however, the rate of these sweeps is low enough not to affect the aggregate spectra (Fig.~\ref{fig_syn}ab). In contrast, the spectra at low recombination rates show a depletion of intermediate and high frequencies. This depletion signals genetic draft on synonymous mutations, which we attribute interference selection. The argument for interference selection will be completed below, where we show that the onset of the shape distortion occurs at a value $\omega^* \sim 1$ predicted by the interference scaling model and is accompanied by a consistent $\rho$-dependence of the synonymous sequence diversity.

To map the transition point between local interference and interference condensate, we calibrate draft model spectra for neutral sites,
\begin{equation}
q_s (x) \approx \theta_s \, Q_0 (x; \nu),
\label{qs}
\end{equation}
to the empirical frequency spectra $\hat q_s^o (x)$ of synonymous sequence sites in {\em D. melanogaster}.
We use a consistent Bayesian inference scheme that includes sampling effects and the map from outgroup-directed to ancestor-directed spectra (Methods). This scheme provides maximum-likelihood values and confidence intervals of the the shape parameter $\nu$ (Fig.~\ref{fig_syn}c) and of the mutation density $\theta_s$ (shown below in Fig.~\ref{fig_omega}a) in each recombination bin, without additional fit parameters. The synonymous spectral data signal an onset of interference selection at a threshold recombination rate $\rho^* \approx \mu$, corresponding to a shape parameter or rescaled draft rate $\nu = 1$.  Regions with $\rho \gtrsim \rho^*$ are inferred to be in the local interference regime, regions with $\rho \lesssim \rho^*$ are in the interference condensate. This regime is characterized by moderate interference with values of the shape parameter in the range $\nu \lesssim 2$. To obtain some insight on the selective effects causing interference, we infer ``corrected'', ancestor-directed sample spectra $\hat q_s (x)$ by an inverse linear map (Methods). The low-recombination spectra $\hat q_s (x)$ are monotonic and well approximated by the spectral functions $Q_0(x; \nu)$ of the draft model, but they do not show the minimum at $x = 1/2$ characteristic of the travelling-wave model (Supplementary Figure~S2). This suggests the underlying interference selection includes drivers under substantial selection and is at some distance from the travelling-wave regime of multiple mutations with individually small effects.

\begin{figure}
  \includegraphics[width=\textwidth]{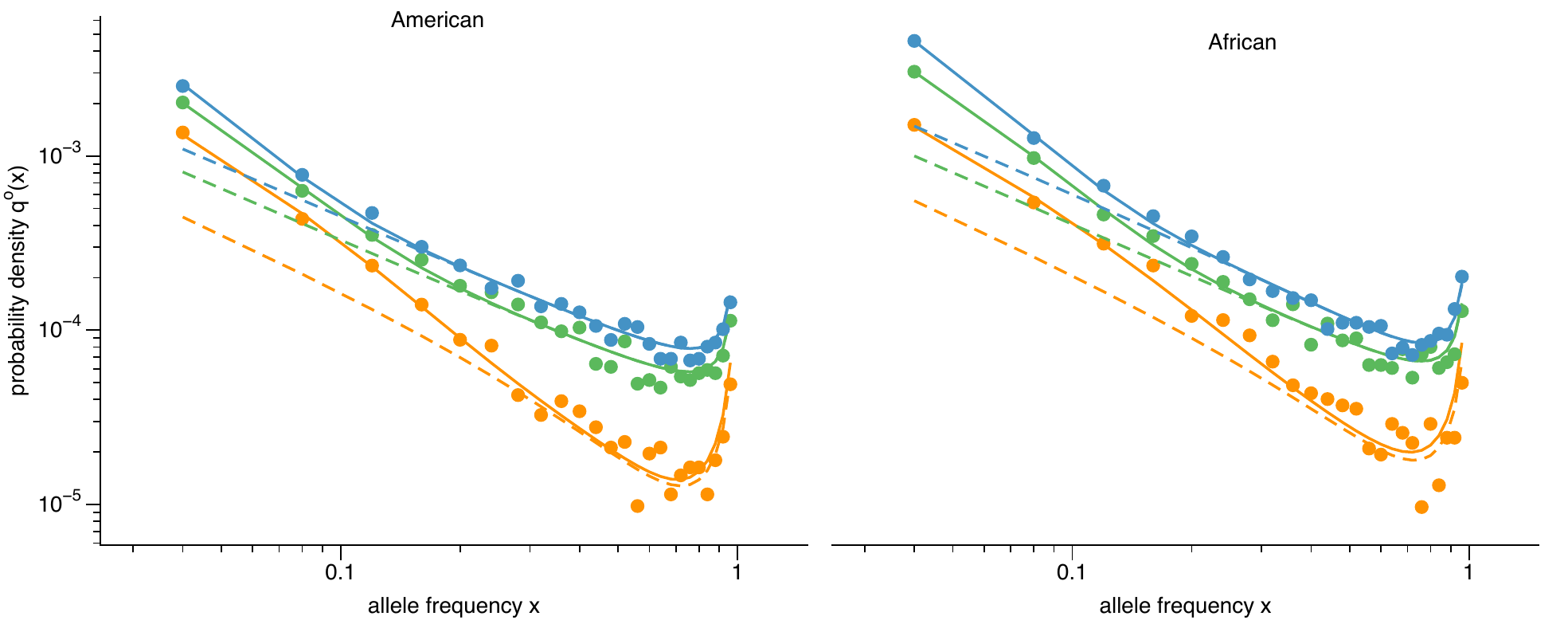}
  \caption{\textbf{Frequency spectra of nonsynonymous mutations.} (a,b) Outgroup-polarized sample spectra $\hat q_a^o(x)$  from two populations of \emph{D. melanogaster}, together with maximum-likelihood spectra $q_a^o(x)$ of the two-component model (solid lines: full spectrum, dashed lines: neutral component). See equations (\ref{qa}) and~(\ref{qao}). Our inference uses the shape parameter $\nu$ inferred from synonymous mutations (Fig.~\ref{fig_syn}c); other inferred model parameters are reported in Supplementary Table~S2.}
  \label{fig_nonsyn}
\end{figure}

The partitioning of the \emph{Drosophila} genome in local interference regions and interference condensate regions can be consistently traced in all sequence categories (Fig.~\ref{fig_nonsyn}, Supplementary Figure S3). At low recombination rates, all of these site frequency spectra show qualitatively the same depletion of intermediate and high frequencies that is characteristic of the condensate regime. Specifically for nonsynonymous mutations, we calibrate a two-component model
\begin{equation}
q_{a} (x) \approx \theta_a \, Q_0 (x; \nu) + \theta'_a \, Q(x; \zeta_a, \nu)
\label{qa}
\end{equation}
to the empirical spectra $\hat q_a (x)$. Here $Q(x; \zeta, \nu)$ is a spectral function for sequence sites with mean selection coefficient $s_a$ that are subject to genetic draft with rate $\sigma$. This function contains branches $Q_\pm (x; \zeta_a, \nu) = {\rm e}^{(\pm \zeta - \nu)x}$ that correspond to beneficial and deleterious mutations, respectively, and depend on the rescaled selection coefficient $\zeta$ (Methods). We obtain maximum-likelihood model parameters $\theta_a, \theta'_a, \zeta_a$ in each recombination bin, using an extended Bayesian inference scheme.  This scheme includes a model for cross-species evolution under selection and the resulting, more complex linear map between ancestor-directed and outgroup-directed spectra (Methods). Remarkably, the spectra for nonsynymous mutations (Fig.~\ref{fig_nonsyn}) can be explained across all recombination classes by the two-component model~(\ref{qa}) with the same shape parameter as inferred for synonymous sites (Fig.~\ref{fig_syn}c) and, hence, with the same threshold rate $\rho^*$. The maximum-likelihood model includes near-neutral sites with spectral shape $Q_0 (x, \nu)$, which is of Kimura form for $\rho > \rho^*$, as well as moderately selected sites with spectrum $Q(x; \zeta_a, \nu)$ and mean effect of order $\zeta_a \sim 20$, which produce excess frequency counts in the range $x \lesssim 0.1$ (see dashed vs. solid lines in Fig.~\ref{fig_nonsyn}). These excess counts cannot be explained by demographic factors, because they are common to both populations and no comparable excess is observed at synonymous sites. Across all recombination rates, the inferred mutation densities $\theta_a$ and $\theta_a'$  are much lower than the density $\theta_s$ at synonymous sites (Fig.~\ref{fig_omega}a), indicating that a large fraction of amino acid changes is under strong selection and hence, suppressed in the frequency range of our spectra. The inferred selective effects of amino acid changes that do appear in our spectra are consistent with the expected fitness landscape of proteins. Important molecular phenotypes of proteins, such as fold stability or enzymatic activity, are quantitative traits encoded by multiple sequence sites. Such traits generically contain weakly and moderately selected constitutive sites, even if the trait itself is under strong stabilizing selection~\cite{Nourmohammad2013-tr}. The maximum-likelihood selection coefficients $\zeta_a$ (Supplementary Table S2) are just one order of magnitude higher than the scaled draft rate $\nu$ in the lowest recombination classes. This suggests that a fraction of nonsynonymous sites is affected by genetic draft in the condensate regime; this point is discussed further below.

\subsection*{Predicting the condensation transition in \emph{Drosophila}}

The threshold recombination rate marking the onset of interference selection is numerically close to the point mutation rate in the \emph{Drosophila} genome, $\mu = 2.8 \times 10^{-9}$ per generation and per unit sequence \cite{Keightley2014-jy} (see Fig.~\ref{fig_syn}b). In light of our scaling theory, this is hardly surprising: the interference density $\omega$, which determines the transition between local interference and interference condensate, is determined by a balance between the rates of local mutations and recombination. We now combine the scaling theory, our evolutionary model, and our Bayesian inference scheme to independently predict the transition point $\rho^*$, as well as the behavior of the sequence diversity in both interference regimes, solely from genomic data in the high-recombination regime. This serves as a stringent consistency test for the scaling theory and provides additional evidence for our inference of interference selection in the \emph{Drosophila} genome.

\begin{figure}
  \includegraphics[width=\textwidth]{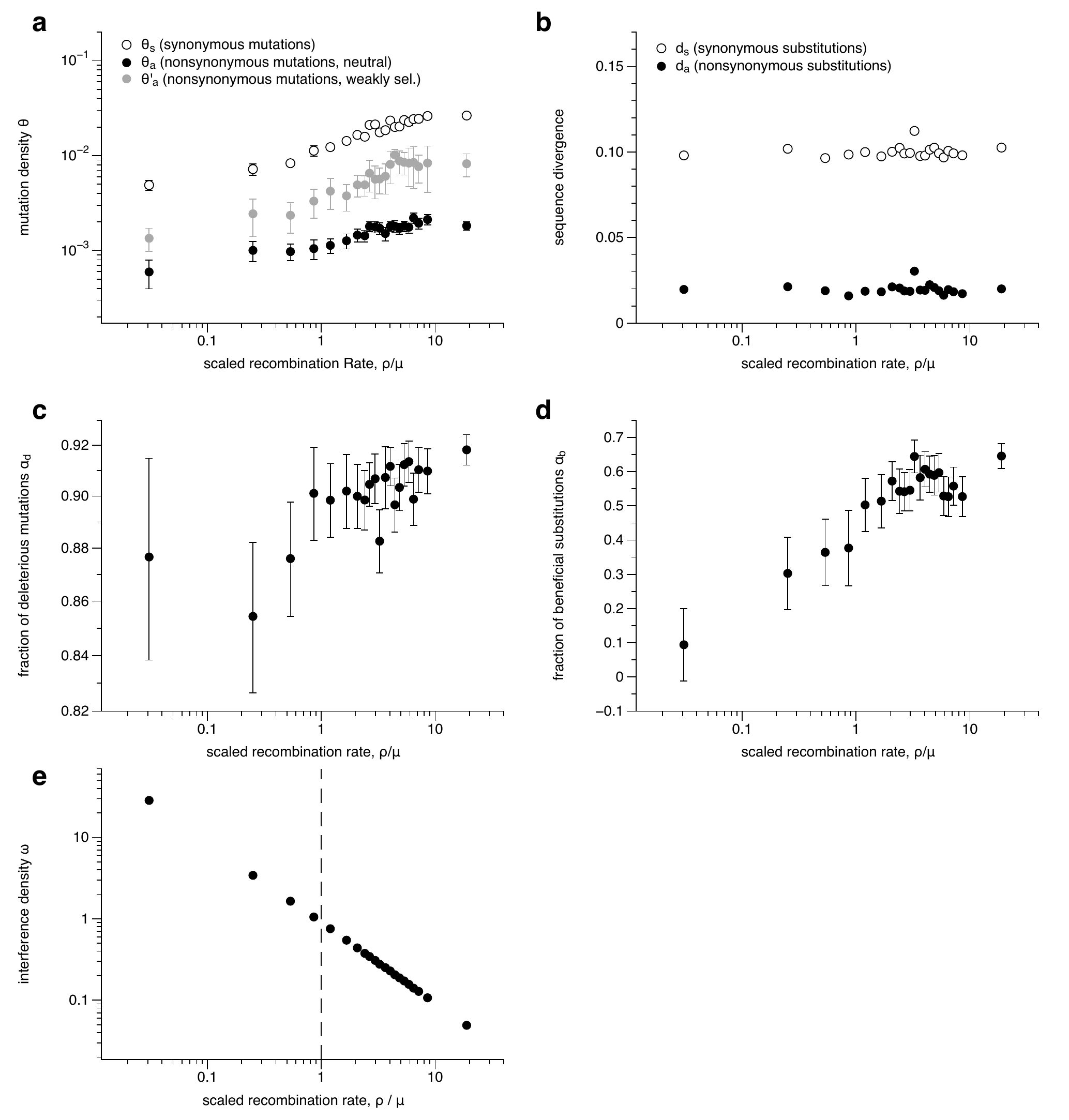}
  \caption{\textbf{Interference density and condensation transition.}
  (a) Mutation densities of synonymous mutations, $\theta_s$, and of nonsynonymous mutations, $\theta_a$ and $\theta'_a$, inferred from spectral data of the African population and plotted against the scaled recombination rate $\rho / \mu$.
  (b) Synonymous divergence, $d_{s}$, and nonsynonymous sequence divergence, $d_a$, to the outgroup species {\em D. simulans}.
  (c) Fraction of deleterious nonsynonymous mutations, $\alpha_d$, as given by equation~(\ref{ud}).
  (d) Fraction of adaptive amino acid substitutions, $\alpha_b$, as given by equation~(\ref{vb}).
  (e)~Interference density, $\omega$, as given by equation~(\ref{omega2}). The condensation transition $(\omega^* \sim 1)$ is mapped to a threshold value $\rho^*/\mu \sim 1$ (dashed line).}
  \label{fig_omega}
\end{figure}

First, we estimate the rate of deleterious mutations in protein-coding sequence from the reduction in mutation density of amino acid changes compared to synonymous changes, $u_d / \mu = \alpha_d$, where $\alpha_d = 1 - (\theta_a /\theta_s)$ is the fraction of amino acid mutations that are deleterious. Moderately deleterious and strongly deleterious amino acid changes contribute partial fractions $\theta'_a /\theta_s$ and $(\theta_s - \theta_a - \theta_a') /\theta_s$, respectively. Second, we estimate the rate of beneficial substitutions in a similar way from the excess of amino acid substitutions compared to the number expected in the near-neutral  component, $v_b / \mu = \alpha_b \,(d_a/d_s)$, where $\alpha_b = 1 - (d_s/d_a)(\theta_a/\theta_s)$ is the fraction of amino acid substitutions that are beneficial. In Methods, we derive these expressions from our evolutionary model and show how they can consistently be extrapolated into the condensate regime. The expression for $\alpha_b$ resembles a McDonald-Kreitman test \cite{McDonald1991-ot,Smith2002-ze}, but our mixed model (\ref{qa}) affords an improved estimate of the mutation density $\theta_a$ by discounting moderately deleterious mutations.
Equation~(\ref{omega}) then gives a simple estimate of the interference density from genomic summary data,
\begin{equation}
\omega = \frac{\mu}{\rho} \left( \alpha_d + \alpha_b \frac{d_a}{d_s }\right) = \frac{\mu}{\rho} \left ( 1 + \frac{d_a}{d_s} - 2 \frac{\theta_a}{\theta_s} \right ).
\label{omega2}
\end{equation}
In Fig.~\ref{fig_omega}ab, we collect the relevant data of the {\em Drosophila} genome in all recombination classes: the maximum-likelihood mutation densities $\theta_s$ and $\theta_a$ inferred from spectral data, and the corresponding sequence divergence levels $d_s$ and $d_a$, defined as the number of substitutions between each {\em D. melanogaster} population and the {\em D. simulans} reference genome. From these data, we infer $\rho$-dependent fractions  $\alpha_d$ and $\alpha_b$ (Fig.~\ref{fig_omega}cd) and the resulting interference density $\omega$ (Fig.~\ref{fig_omega}e).

For $\rho > \rho^*$, we consistently find a deleterious mutation rate $u_d \approx 0.9 \mu$ and a beneficial substitution rate $v_b \approx 0.1 \mu$ that are approximately independent of $\rho$ (Fig.~\ref{fig_omega}cd). Hence, the local interference regime has an approximately constant density of interference domains and an interference density that is inversely proportional to the recombination rate, $\omega \sim 1.0 \, \mu /\rho$. As inferred above, the {\em Drosophila} genome includes a sizeable fraction of moderately selected sites and genome-wide positive selection is not dominated solely by strong selective sweeps; these characteristics suggest the minimal scaling theory in terms of the interference density is applicable. This theory makes three quantitative predictions:
\begin{enumerate}[(a)]
\item
In the local interference regime, the mutation densities $\theta_s$ and $\theta_a$, as well as the sequence diversity are proportional to the probability of no interference,
$p_0 = {\rm e}^{-\omega} \approx {\rm e}^{-\mu /\rho}$. This formula generalizes the standard model of background selection, which predicts the size of the error-free sequence class to depend exponentially on the rate of deleterious mutations~\cite{Charlesworth1993-nk}. Indeed, the observed $\rho$-dependence of the sequence diversity at synonymous sites is in good agreement with this theory in the local interference regime, $\pi_s \sim {\rm e}^{-\mu/\rho}$ (Fig.~\ref{fig_syn}d, see definition in Methods), in broad agreement with previous observations \cite{Begun1992-fe,Mackay2012-yt}.

\item The transition to the interference condensate occurs at a threshold interference density $\omega^*$ of order 1. This determines the threshold recombination rate $\rho^* \sim \mu$ (Fig.~\ref{fig_omega}e), in agreement with the observed onset of interference selection (Fig.~\ref{fig_syn}c).

\item In the interference condensate, the sequence diversity depends only weakly on the recombination rate. This dependence can be derived from a simple scaling argument based on extremal value statistics: an expected number $\omega / \omega^*$ of beneficial mutations with average selection coefficient $\bar s$ originate in each interference domain, but only the fittest of these mutants reaches fixation. This determines the draft rate
$\sigma \approx \bar s (1 + \log (\omega/\omega^*))$, which sets the sequence diversity $\pi_s \simeq 2\mu / \sigma$~\cite{Gillespie2000-fa,Desai2013-nl,Neher2013-qj}. With condensate interference densities bounded in the range $\omega \approx (0.85 - 1.0) \mu /\rho$ (Fig.~\ref{fig_omega}e), we obtain the leading $\rho$-dependence $\pi_s \sim (1 + \log (\rho^*/\rho))^{-1}$, which is in agreement with the observed pattern (Fig.~\ref{fig_syn}d). Our scaling argument is consistent with the scaling of $\sigma$ in the numerical simulations (Fig.~\ref{fig_sim}c)  and with previous results for evolution solely under beneficial mutations \cite{Weissman2012-of}. In a state of stationary fitness, however, beneficial substitutions are a generic feature of the condensate regime. Even in the absence of adaptation, they compensate the fixation of deleterious mutations fixed by interference selection \cite{Mustonen2009-st,Rice2015-qq}. Below, we will discuss the likely loci of these dynamics in the \emph{Drosophila} genome.
\end{enumerate}

\begin{figure}
  \includegraphics[width=\textwidth]{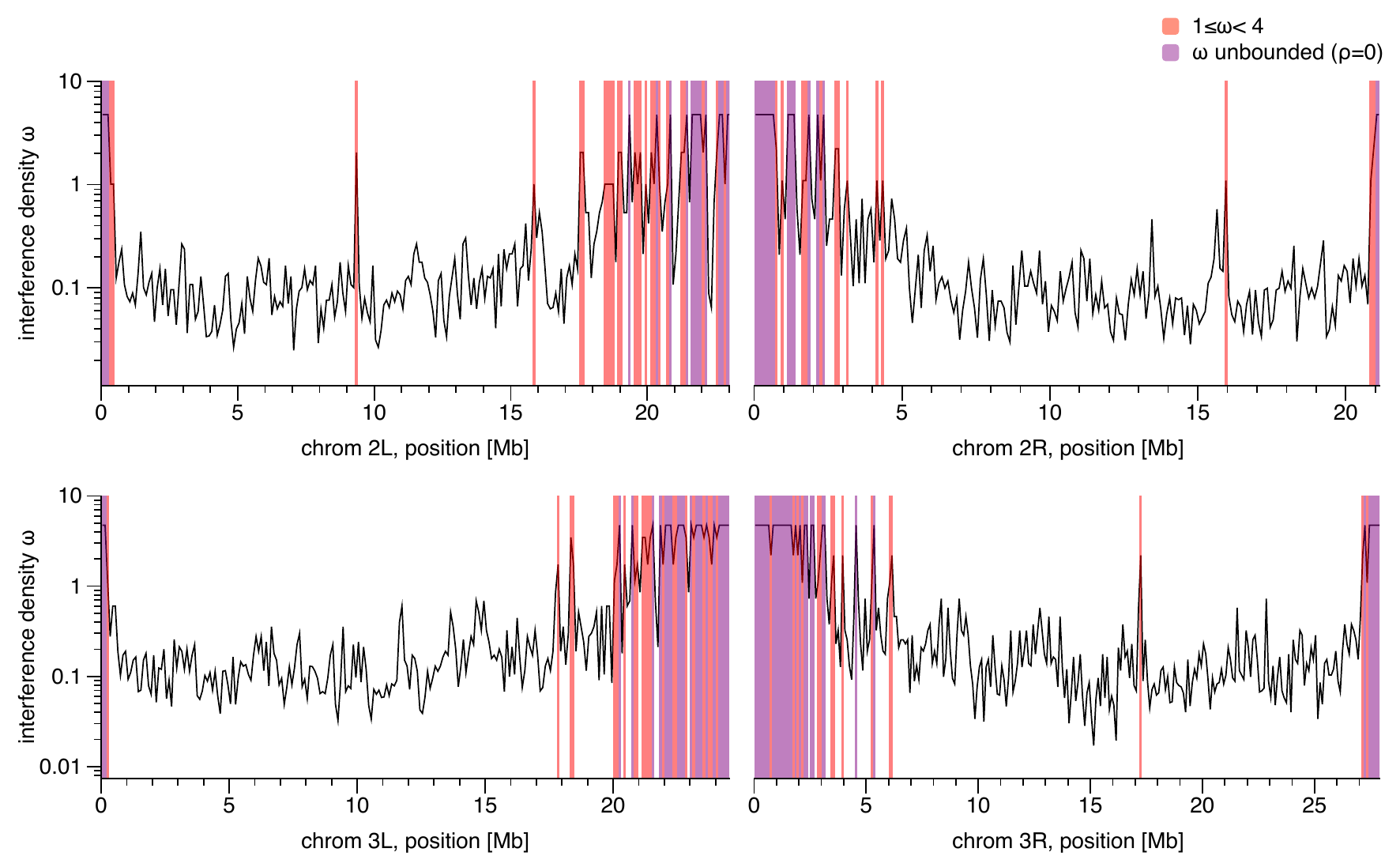}
  \caption{\textbf{Map of evolutionary modes in the \emph{Drosophila} genome.} Position-dependent local interference density $\omega$ in the four autosomes, inferred from the recombination rate $\rho$ in windows of 100kb (see text and Fig.~\ref{fig_omega}e). Condensate segments ($\omega > 1$) cover about 20\% of the genome and are located primarily in chromosomal regions flanking the centromeres (orange: segments with $1 < \omega < 4$; violet: segments with no recorded recombination, $\omega > 4$). We use the map of local recombination rates from ref. \cite{Comeron2012-jh}.}
  \label{fig_map}
\end{figure}

Taken together, genetic variation in \emph{Drosophila} is in remarkable quantitive agreement with our interference scaling theory over two decades of recombination rates. Fig.~\ref{fig_map} charts the local interference density $\omega$ in the autosomes of \emph{D. melanogaster}, the high-resolution recombination map of ref.~\cite{Comeron2012-jh} and our genomic inference. Extended condensate regions, shown in orange, are located primarily adjacent to the centromere regions and, to a lesser extent, to the telomeres. The major part of condensate sequence maintains a residual level of recombination, corresponding to interference densities in the range $1 < \omega \leq 4$. The remaining 9\% of the autosomal genome consists of 38 contiguous segments with no recorded recombination ($\omega > 4$); these segments  have an average length of 0.2 Mb and a maximum length of 0.9Mb. We now turn to inferring key genomic and evolutionary features of the interference condensate.

\subsection*{Linkage correlations in the condensate}

\begin{figure}
  \includegraphics[width=\textwidth]{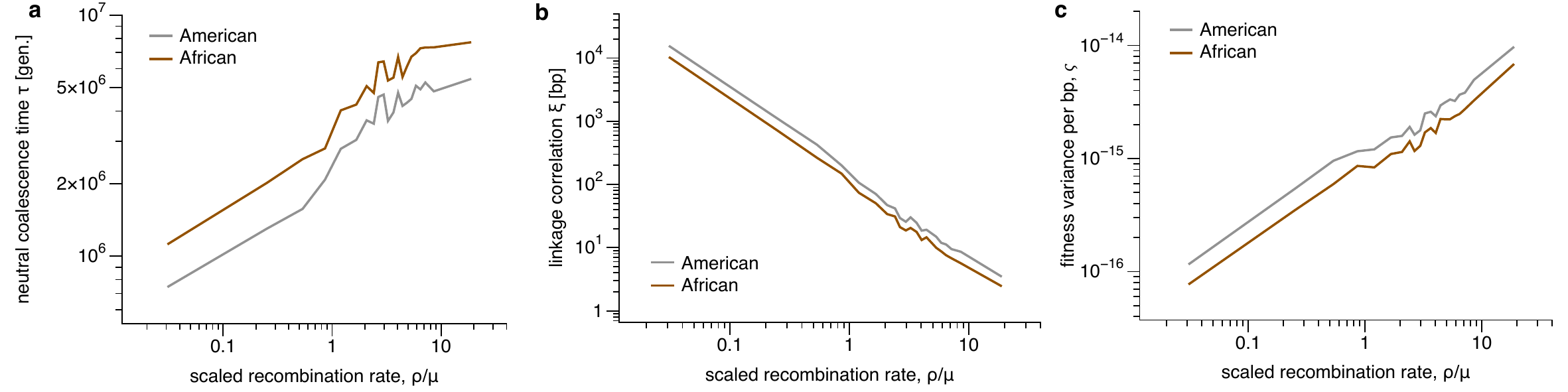}
  \caption{\textbf{Interference domains and speed of evolution in the condensate.} (a) Coalescence time, $\tau = 1/\tilde \sigma$, and (b) linkage correlation length, $\xi$, plotted against the scaled recombination rate $\rho/\mu$. In the condensate regime, these scales determine the characteristic shape of interference domains (Fig.~\ref{fig_schematic}b). See equations~(\ref{tauxi1}) and~(\ref{pi2}). (c)~Fitness variance per site, as given by equation~(\ref{speed}). This quantity displays a drastically reduced efficacy of selection in the condensate, consistent with the reduced fraction of adaptive substitutions (Fig.~\ref{fig_omega}d).}
  \label{fig_xi}
\end{figure}

Although the condensate is a complicated regime of strongly correlated mutations, it has remarkably simple emergent scaling properties. Because interference domains in the condensate are densely packed, the draft rate  $\sigma$ becomes similar to the neutral coalescence rate, $\tilde \sigma$, which is also the scale of fitness differences between competing clades. The emergence of a characteristic scale of genetic turnover is a common feature of models of asexual evolution \cite{Schiffels2011-wn,Neher2013-hs}. Under finite recombination, the rate $\tilde \sigma$ sets the genomic correlation length $\xi = \tilde \sigma/\rho$; coexisting mutations at a distance $r \lesssim \xi$ are likely to retain their genetic linkage over a mean coalescence time interval $\tau = 1/\tilde \sigma$ (Fig.~\ref{fig_schematic}b). We can estimate the coalescence rate from the inferred values of draft rate and synonymous mutation density~\cite{Schiffels2011-wn}, $\tilde \sigma = \sigma + 2 \mu/\theta_s$, or equivalently from the neutral sequence diversity $\pi = 2\mu / \tilde \sigma$ (Methods). Together, we obtain the simple estimates
\begin{equation}
\tilde \sigma = \frac{2\mu}{\pi},
\qquad
\xi = \frac{2\mu}{\rho \pi},
\label{tauxi1}
\end{equation}
which determine the universal shape of interference domains in the condensate regime (Fig.~\ref{fig_schematic}b). In coalescent models under selection, the same scaling $\tilde \sigma \sim \mu / \pi$ links the coalescence rate to the neutral sequence diversity, in some cases with logarithmic corrections \cite{Desai2013-nl,Neher2013-qj}. In the condensate regime, we find coalescence times $\tau$ about an order of magnitude lower than at high recombination rates  (Fig.~\ref{fig_xi}a) and genomic correlations up to $\xi \lesssim 10^4$ base pairs (Fig.~\ref{fig_xi}b), signalling that neighboring genes are often in common interference domains.

\subsection*{Speed and cost of evolution in the condensate}

The most important asexual feature of the interference condensate is the drastically reduced efficacy of selection. In the \emph{Drosophila} genome, we can quantify this effect in two ways. First, the fraction of beneficial substitutions, $\alpha_b$, which takes stable values of about $50\%$ in the local interference regime, sharply drops in the condensate to below $10\%$ in the lowest recombination class (Fig.~\ref{fig_omega}d). Second, the fitness variance per unit sequence of the condensate is related to the neutral sequence diversity,
\begin{equation}
\varsigma^2  = \frac{2\rho \mu}{\pi}
\label{speed}
\end{equation}
(Methods). The $\rho$-dependent values of $\varsigma^2$ inferred from the synonymous sequence diversity $\pi_s$ (Fig.~\ref{fig_xi}c) show a sharp drop in the efficacy of selection within the condensate regime; the fitness variance in the lowest recombination class is by a factor 10 lower than at the transition point $\rho^*$. The strong dependence of the fitness variance on the recombination rate is in tune with the simulation results (Fig.~\ref{fig_sim}a). We conclude that interference selection curbs rate and selective effects of adaptive evolution in the condensate regions of the \emph{Drosophila} genome.

The reduced efficacy of selection has an immediate consequence for genome functionality in the condensate regime: interference selection generates \emph{emergent neutrality} of sequence sites with selection coefficients $s \lesssim \sigma$; these sites become disfunctional because their alleles are randomized by interference selection \cite{Schiffels2011-wn}. We can estimate the resulting fitness cost (or genetic load) for a protein, $\Delta F = (\ell/2) \int_0^\sigma s \, f (s) \, ds$, where $f (s)$ is the distribution of selection coefficients and $\ell$ is the length of the protein. This cost increases with decreasing $\rho$, because $\sigma$ as inferred through the shape parameter increases (Fig.~\ref{fig_syn}c). Emergent neutrality says that the genetic load in the two modes differs not only in magnitude, but qualitatively. In the local interference regime, a genetic locus under moderate selection ($s > 1/2N$) incurs classical mutational load, where the beneficial allele is always prevalent and only a small minority fraction of the population, of average $\mu/s$, carries the deleterious allele. In the condensate, deleterious and compensatory beneficial substitutions generate a new equilibrium in which the deleterious allele becomes dominant in the population with probability $1 / (1 + {\rm e}^{s/\sigma})$ \cite{Schiffels2011-wn}. Hence, the functional impact of moderately deleterious alleles ($s < \sigma$) becomes important. Emergent neutrality likely affects part of the nonsynonymous sites in the moderate selection class (Fig.~\ref{fig_nonsyn}), as well as intron, UTR, and synonymous sites under selection for codon usage. Assuming that just a few percent of these sites become effectively neutral, the above estimate predicts a substantial scaled fitness cost $2 N \Delta F \sim 10-100$ per gene, even if the effect of each individual site is weak. This fitness cost is specific to genes in the condensate regime; its likely consequences for genome architecture are discussed below.

\section*{Discussion}

The main method development of this paper is a unified scaling theory of genetic draft and background selection. This theory identifies a dominant scaling variable, the interference density $\omega = (u_d + v_b)/\rho$, to discriminate between two evolutionary modes: the local interference regime ($\omega \lesssim \omega^*$) and the interference condensate ($\omega \gtrsim \omega^*$). In the local interference mode, mutations evolve in an approximately independent way by selection and genetic drift; in the condensate, they are locked into clades of genetically linked sequence segments and many are governed by linkage. This mode requires a sufficiently high supply of mutations under substantial (beneficial or deleterious) selection but is insensitive to details of the evolutionary process -- in particular, to the rate of adaptation. Over a broad range of evolutionary parameters, the transition point $\omega^*$ between local interference and condensate is of order 1. The frequency spectrum of neutral mutations can be used as a marker of the evolutionary mode: the ``convective'' frequency evolution in the condensate regime is signalled by a characteristic depletion of intermediate and high frequencies.

In the \emph{Drosophila} genome, we build a case for the joint presence of these evolutionary modes from a number of mutually consistent observations from sequence data. We infer the rates of deleterious amino acid changes, $u_d$, and the rate of beneficial substitutions, $v_b$, in protein coding sequence. Given these selective building blocks of the interference density $\omega$, our scaling theory predicts how genetic variation depends on recombination: the sequence diversity varies strongly in the local interference regime, $\pi \sim {\rm e}^{- \omega}$, and weakly in the condensate, $\pi \sim (1 + \log (\omega/\omega^*))^{-1}$; the transition point between these regimes, $\omega^* \sim 1$, marks the onset of genetic draft. Together, amplitude and shape of mutational spectra change in a concerted way. These predictions are in agreement with direct genomic data of synonymous mutations. While any single characteristic of genetic variation could be explained by alternative evolutionary scenarios, the consistent joint pattern of diversity and spectral shape over the entire range of recombination rates provides strong evidence for an interference condensate in the \emph{Drosophila} genome (Fig.~\ref{fig_syn}cd).

Our results suggest that the established rationale of a strong evolutionary advantage of sex applies to about 80\% of the Drosophila genes, which are in the local interference regime. The other 20\%, some 3000 genes in the interference condensate, show evolutionary similarities with asexual systems. In the condensate regions, we infer a significantly lower fitness variance per unit sequence, indicating reduced evolvability in response to adaptive pressure. This may signal that condensate genes respond less efficiently to existing positive selection for change or that they are subject to less selection for change in the first place. We also infer a significantly increased fitness cost (genetic load) concentrated in weakly and moderately selected sequence sites, whose alleles are randomized by emergent neutrality \cite{Schiffels2011-wn}. This finding suggests that the evolutionary partitioning of the \emph{Drosophila} genome is also a functional partitioning. We hypothesize that condensate genes have systematically lower intrinsic fold stability than other genes. They should also have reduced codon usage bias, which may affect speed and efficiency of translation and, hence, increase the cost of protein expression. These hypotheses on the functional impact of interference selection can be tested by experiment and by targeted sequence analysis.

A salient feature of \emph{Drosophila} is that both evolutionary modes coexist in one genome. This implies that functional and fitness differences between condensate genes and other genes play out in the same individual, the same environment, and the same population. Over macro-evolutionary time scales, these differences can generate feedback effects on genome architecture. First, we expect selection against too long recombination coldspots. This is qualitatively in line with observations: in 91\% of the autosomes, the \emph{Drosophila} genome maintains a residual level of recombination, keeping interference selection capped to a moderate level ($\omega \leq 4$); the remaining zero-recombination sequence is fragmented into short contiguous segments (Fig.~\ref{fig_map}). Second, a given gene incurs a fitness cost that depends on its target function and on the interference regime it is placed in. Therefore, genes with high requirements on protein stability or translation efficiency should be suppressed in the condensate. Whether there are differences in gene content and gene functions between the local interference regime and the condensate that can be explained as a consequence of differences in interference selection is an interesting question for future research.

\subsection*{Methods}

\subsubsection*{Scaling theory}
The heuristic scaling approach used in this paper is based on three main ingredients: (i) In the local interference regime, the behavior of an evolutionary observable can be calculated approximately from single-site population genetics. (ii) The crossover to the interference condensate regime can be described by a scaling function that depends only on the variable $\omega$ given by equation~(\ref{omega}). Here and in the following, crossover is used as a technical term of scaling theory that is not to be confused with the genetics term. (iii) In the condensate, evolutionary observables follow broad heuristic constraints, and there is a matching condition between both scaling regimes at the crossover point $\omega^*$. We consider long genomic segments that evolve under limited recombination; individual sequence sites have a distribution of selection coefficients with $2N \bar s \gg 1$, where $\bar s$ is the average selection coefficient and $N$ is the effective population size.
For simplicity,
we neglect prefactors of order 1 and corrections to scaling, which often depend on more specific model assumptions. Our analysis in the main text builds on a minimal model with the following scaling relations:
\begin{enumerate}[(a)]
\item The average fitness variance per unit sequence, $\varsigma^2$, takes the form
\begin{equation}
\varsigma^2 \simeq \left \{
\begin{array}{ll}
(u_d + v_b) \bar s & (\omega \lesssim 1, \mbox{local interference})
\\ \\
\rho \sigma & (\omega \gtrsim 1, \mbox{interference condensate}),
\end{array} \right.
\label{varsigma}
\end{equation}
where $\sigma$ is a characteristic selection strength in the condensate regime. The
local interference expression $\varsigma^2 \simeq (u_d + v_b)\bar  s$ follows from direct calculation by single-site population genetics, assuming statistical independence of selected alleles at different sites. The leading condensate asymptotics $\varsigma^2 \simeq \rho \sigma$ is then already determined by the scaling properties (ii) and (iii). Specifically, we evaluate the matching condition $\sigma (\omega^*) = \bar s$ at the crossover point $\omega^* \sim 1$ of the minimal scaling theory with the requirement that $\varsigma^2$ depends only weakly on $u_d$ and $v_b$ in the condensate. This is expected from the jamming of genomic space-time shown in Fig.~\ref{fig_schematic}b and implies that the recombination rate becomes a limiting factor of $\varsigma^2$ in the condensate. The scaling argument given in the main text suggests a specific functional form in the condensate, which is given by
\begin{equation}
    \sigma \simeq \bar s \, (1 + \log (\omega/\omega^*));
    \label{eq:sigma_condensate}
\end{equation}
see also ref.~\cite{Weissman2012-of}. Equation~(\ref{varsigma}) can be rescaled to a dimensionless form,
\begin{equation}
\frac{\varsigma^2}{\rho s} \simeq \left \{
\begin{array}{ll}
\omega & (\omega \lesssim 1, \mbox{local interference})
\\ \\
\dfrac{\sigma}{s} & (\omega \gtrsim 1, \mbox{interference condensate}),
\end{array} \right.
\label{varsigma2}
\end{equation}
which is confirmed by our simulation results (Fig.~\ref{fig_schematic}c). Equations (\ref{varsigma}) and (\ref{varsigma2}) and, in particular, the minimal crossover scaling $\omega^* \sim 1$ are consistent with previous results for evolution under beneficial mutations~\cite{Weissman2012-of} and under background selection including moderate effects ($\log (2N s) \sim 1$) \cite{Hudson1995-kx,Nordborg1996-xc}. Strong heterogeneities in the effect distribution or background selection by strongly deleterious mutations generate systematic shifts of the crossover point $\omega^*$; the corresponding extensions of scaling theory are discussed below.

\item As explained in the main text, $\sigma$ determines the characteristic scales of time and genomic distance in the condensate\cite{Weissman2012-of},
\begin{equation}
\tau \simeq \frac{1}{\sigma},
\qquad
\xi \simeq \frac{\sigma}{\rho}
\qquad (\omega \gtrsim 1, \mbox{ interference condensate}).
\label{tauxi}
\end{equation}

\item The neutral sequence diversity, $\pi = 2 \int_0^1 x (1-x) \, q (x) \, dx$, can be estimated by the form
\begin{equation}
\pi \simeq \left \{
\begin{array}{ll}
2 \mu N {\rm e}^{-\omega} & (\omega \lesssim 1,  
\mbox{ local interference})
\\ \\
\dfrac{2\mu}{ \sigma} & (\omega \gtrsim 1, \mbox{ interference condensate}).
\end{array} \right.
\label{pi}
\end{equation}
Using again the matching criterion $\sigma (\omega^*) = \bar s$ at the crossover point $\omega^*$, equation (\ref{pi}) interpolates between background selection in the local interference regime and neutral evolution under genetic draft with rate $\sigma$ in the condensate regime \cite{Hudson1995-kx,Nordborg1996-xc,Good2014-bq,Neher2013-qj}. In the local interference regime, $\pi$ is proportional to~$N$, which implies that established neutral mutations evolve predominantly under genetic drift and their spectrum is of Kimura form. Hence, the genetic draft of neutral mutations and the resulting depletion of intermediate and high frequencies is a marker of the condensate regime, which sets on at the crossover point $\omega^*$. Equation~(\ref{eq:sigma_condensate}) then predicts the form of the diversity in the condensate regime,
\begin{equation}
    \pi \simeq \frac{2\mu}{ \bar s (1 + \log (\omega/\omega^*))}.
    \label{eq:pi_condensate}
\end{equation}
\end{enumerate}
Equations~(\ref{varsigma}) -- (\ref{pi}) show that $\pi$ measures key characteristics of the condensate regime, the spacetime scaling (equation~(\ref{tauxi1})) and the fitness variance per unit sequence (equation~(\ref{speed})). In the main text, we use the sample sequence diversity at synonymous sites, $\pi_s = 2 \sum_{k=0}^{n} (k/n) (1-k/n) \, \hat q^o_s (k/n)$ to infer these characteristics in the {\em Drosophila} genome. As discussed in the main text, the {\em Drosophila} genome has a distribution of selective effects that includes sites with weak and moderate selection, for which the minimal scaling theory with $\omega^* \sim 1$ should be applicable. We find clear evidence that $\pi_s$ follows the scaling behavior predicted by equations (\ref{pi}) and~(\ref{eq:pi_condensate}); see Fig.~\ref{fig_syn}d.

\subsubsection*{Extensions of scaling theory}
Evolution by beneficial and deleterious mutations is a complex process whose details depend on their rates and effect distribution. The minimal scaling theory is a coarse approximation of this process. It provides useful approximations of genomic statistics over a wide range of evolutionary parameters, which includes settings appropriate for \emph{Drosophila}. Here we discuss two extensions that serve to link our scaling theory to existing evolutionary models and to delineate the range of validity of the minimal model.

\begin{enumerate}[(a)]
\item Background selection with strong effects. Equation~(\ref{pi}) predicts the onset of interference selection for sites of selection coeffcient $s$ at a characteristic value
\begin{equation}
\omega^* = \log ( Ns).
\label{match}
\end{equation}
This expression is consistent with known results of background selection theory \cite{Charlesworth1994-lm,Good2014-bq,Neher2013-qj,Jain2008-js}. If background selection involves only strongly selected sites ($2N s \gg 1$), we obtain a shift of the crossover point $\omega^*$ observed in aggregate data to values above 1 (Supplementary Figure~S1a). The crossover point is still marked by the onset of interference selection on neutral sites and the resulting spectral shape distortion. This regime is not relevant for \emph{Drosophila}, where we observe $\omega^* \sim 1$ and consistently, genomic sites under weak and moderate selection.

\item Selective sweeps. An adaptive process driven by strongly beneficial mutations with rate $v_b$ and average selection coefficient $s_b$ generates a fitness flux $\phi = v_b s$, which measures the speed of adaptation per unit sequence \cite{Mustonen2007-kq}. Under this process, a genomic focal site is subject to linked sweeps at a rate $\sigma_{\rm sweep} = v_b \xi_s = v_b s_b /\rho = 2N \phi / \rho$. Focal sites of selection coefficient $s < s_b$ are strongly affected by interference for $\sigma_{\rm sweep} > s$, which sets the crossover point to interference selection,
\begin{equation}
\omega^* = \frac{s}{s_b} \leq 1.
\end{equation}
In the special case of equal selection coefficients at all sites ($s = s_b$), the crossover point is again $\omega^* = 1$, independently of $s_b$~\cite{Weissman2012-of,Good2012-ne} (Supplementary Figure~S1b).
The onset of interference on neutral sites and the resulting spectral shape distortion occur at a value $\omega_0 = 1/(2Ns_b) = \omega^* / (2Ns)$, which is smaller than $\omega^*$. This regime is not observed in  \emph{Drosophila}; strong sweeps are too rare to distort neutral spectra for $\omega < 1$.
\end{enumerate}

\subsubsection*{Link to asexual evolution}
An explicit expression for $\sigma$ can be obtained if we identify the total fitness variance per correlation interval, $\xi \varsigma^2 = \sigma^2$, with the corresponding quantity in models of asexual populations with a genome of length $\xi$ \cite{Weissman2012-of}. In the intermediate fluctuation regime of travelling waves, this identification yields the analytic expression $\sigma = \bar s \log (\dots)$, where $\log (\dots)$ stands for a weak and model-specific dependence on the parameters $u_b$, $\bar s$, $N$ and the system size $\xi$; see equation (15) of ref. \cite{Good2012-ne}. This form has a leading large-$\xi$ asymptotics consistent with our scaling argument given in the main text, $\sigma \simeq \bar s \log \xi \simeq \bar s \log (\rho^*/\rho)$. In ref.~\cite{Neher2013-qj}, an analogous identification is discussed for evolutionary processes dominated by weakly selected alleles.

The theoretical limit of strictly asexual evolution, which is reached at very low recombination rates $\rho \ll \sigma/L$ in a genome of length $L$, can be described in terms of our scaling theory by substituting $L$ for the correlation length $\xi = \sigma/ \rho$ resp. $\xi_s = s/\rho$. In this limit, the interference density (\ref{omega}) for mutations of effect $s$ takes the form
\begin{equation}
\omega = \frac{L (u_d + v_b)}{s} = \frac{U_d + V_b}{s},
\end{equation}
which depends on the genome-wide rates  $U_d = L u_d$ and $V_b  = L v_b$. For $V_b = 0$, the identity~(\ref{match}) for $\omega^*$  becomes the well-known criterion for the onset of Muller's ratchet in asexual populations, $U_d / s \sim \log ( N s)$  \cite{Muller1964-vh,Felsenstein1974-sr,Jain2008-js,Goyal2012-ro}.

\subsubsection*{Evolutionary model: mutation frequency distributions}

As explained in the main text, we use a specific model of interference selection to parametrize site frequency spectra:  individual sites evolve under mutations (with rate $\mu$), selection (with site selection coefficient~$s >0$), genetic drift (in a population of effective size $N$), and periodically recurrent genetic draft (with rate $\sigma$). The draft model generates site frequency spectra that can be estimated analytically by a saddle-point approximation to the path integral of mutation frequency paths \cite{Mustonen2010-oz}. For beneficial mutations (of selection coefficient $s$) and deleterious mutations (of selection coefficient $-s$) at two-allelic sites, we obtain the frequency distributions
\begin{eqnarray}
q_\pm (x) & =  & \frac{1}{Z_\pm (\theta, \zeta, \nu)} \, q_0 (x) \,
\int p (\tau(\tilde s, x), \sigma) \, \exp \left [ - \frac{\theta_0}{4N\theta}  \int_0^{\tau(\tilde s, x)} \frac{\big (\dot x(t; \tilde s) \mp s g(x(t; \tilde s)) \big)^2 }{g(x(t; \tilde s))} \, {\rm d}t \right ] \, d\tilde s
\nonumber
\\
\label{qsel1}
\\
& \simeq & \theta \, Q_\pm (x; \zeta, \nu)
\qquad \Big ( \frac{1}{N} \leq x \leq 1 - \frac{1}{N} {\Big)},
\label{qsel}
\end{eqnarray}
respectively; these distributions depend on the mutation density $\theta = \theta_0 {\rm e}^{-\omega} = \mu N {\rm e}^{-\omega} \ll 1$, the scaled draft rate $\nu = 2N \sigma \theta/\theta_0$, and the scaled site selection coefficient $\zeta = 2N s  \theta/\theta_0$. The function $q_0 (x) = x^{-1 + \theta} (1-x)^{-1 + \theta}$ denotes the neutral spectrum under genetic drift. The no-sweep probability $p(\tau, \sigma)$ over a time interval $\tau$ is assumed to be strongly suppressed for $\tau \gtrsim  1/\sigma$. The exponential weight involves the maximum-likelihood frequency path with effective selection coefficient $\tilde s$, which is denoted by $x(t, \tilde s)$. This path follows the equation of motion $\dot x(t, \tilde s) = \tilde s g (x(t, \tilde s))$ with $g(x) = x (1-x)$ and has a sojourn time $\tau(\tilde s, x)$ up to frequency $x$. The prefactor $Z_\pm (\theta, \zeta, \nu)$ ensures the normalization $\int_0^1 q (x) \, dx = 1$. An approximate evaluation of the integral in equation~(\ref{qsel1}) results in the remarkably simple spectral function
\begin{equation}
Q_\pm (x; \zeta, \nu) = \frac{{\rm e}^{ (\pm \zeta - \nu) x}}{x}.
\label{Q}
\end{equation}
This function consistently interpolates between the asymptotic regimes of effectively neutral mutations ($\zeta \ll \nu$, i.e., $\tilde s \simeq \sigma$), which are dominated by genetic draft ($Q \simeq {\rm e}^{- \nu x}/x$), and strongly selected mutations ($\zeta \gg \nu$, i.e., $\tilde s \simeq s$), which evolve in an autonomous way ($Q \simeq {\rm e}^{\pm \zeta x}/x$). For the spectral function of neutral sites, we use the shorthand
\begin{equation}
Q_0 (x; \nu) = \frac{{\rm e}^{ - \nu x}}{x}.
\label{Q0}
\end{equation}
The family of spectral functions (\ref{Q}) provides a good parametrization of the spectral data in our simulations (Fig.~\ref{fig_schematic}d), as well as in all sequence classes and recombination rate classes of {\em Drosophila} (Fig.~\ref{fig_syn}ab, Fig.~\ref{fig_nonsyn}, and Supplementary Figure~S3). Frequency distributions of the form (\ref{qsel}) serve as building blocks for our genomic inference; see equations (\ref{qs}) and (\ref{qa}) of the main text and equations (\ref{qso}) -- (\ref{qao}) below.

The spectral functions of the draft model map the crossover from drift-dominated to draft-dominated evolution in analytical form. The neutral site frequency spectrum (\ref{qs}) determines the sequence diversity
\begin{equation}
\pi \approx \frac{\theta}{1 + \nu /2},
\label{pi1}
\end{equation}
which is consistent with the scaling behavior~(\ref{pi}). In the local interference  regime $\pi \simeq \theta = \mu N {\rm e}^{-\omega}$; i.e., background selection reduces diversity but does not affect the shape of the frequency spectrum. In the condensate regime, $\pi \simeq 2 \mu/ \sigma < \theta$, i.e., diversity and spectral shape are determined by interference selection~\cite{Gillespie2000-fa}. These features hold for broad classes of interference selection \cite{Neher2013-xf}, making the spectral functions a convenient choice for parametrizing the {\em Drosophila} site frequency spectra. Specifically, we use the condition $\nu > 1$ on the shape parameter inferred from the spectrum of synonymous sequence sites as a marker of the interference condensate regime.

\subsubsection*{Evolutionary model: substitution dynamics and allele occupancy}

The draft model also serves to parametrize the sequence evolution between the ingroup species {\em D. melanogaster} and the outgroup species {\em D. simulans}. In this model, allele substitutions at individual sites take place with Kimura-Ohta rates that depend on the local coalescence rate (or inverse effective population size)
\begin{equation}
\tilde \sigma = \frac{2\mu}{\pi} = \sigma + \frac{2 \mu}{\theta} \simeq \left \{
\begin{array}{ll}
\dfrac{{\rm e}^{\omega}}{2N} & (\omega \lesssim 1,  \mbox{ local interference})
\\ \\
\sigma & (\omega \gtrsim 1, \mbox{ interference condensate}).
\end{array} \right.
\label{pi2}
\end{equation}
Consistently with equations~(\ref{pi}) and~(\ref{pi1}), the coalescence rate maps again the crossover between local interference and condensate regime. The beneficial and deleterious substitution rates
\begin{equation}
u_\pm (\zeta, \nu) = \mu \, \frac{\zeta/\tilde \nu}{1 - {\rm e}^{-\zeta / \tilde \nu }},
\label{upm}
\end{equation}
depend on the scaled selection coefficient $\zeta$  and the scaled coalescence rate $\tilde \nu \equiv 2N \tilde \sigma \theta/\theta_0 = \nu + 1$. Models of this form have been shown to provide an excellent approximation to the equilibrium substitution dynamics in linked genomes under different scenarios of interference selection \cite{Schiffels2011-wn,Rice2015-qq}. The rates (\ref{upm}) consistently determine the equilibrium occupancy probabilities of beneficial and deleterious alleles,
\begin{equation}
\lambda_\pm (\zeta, \nu) = \frac{1}{1+ {\rm e}^{ \mp \zeta / \tilde \nu }},
\label{lambdapm}
\end{equation}
as well as the expected sequence divergence between in- and outgroup species,
\begin{equation}
d (\zeta, \nu) = \tau \big [\lambda_- (\zeta, \nu) u_+ (\zeta, \nu) + \lambda_+ (\zeta, \nu) u_- (\zeta, \nu) \big] = \frac{d_0}{(\zeta/\tilde \nu) \sinh (\zeta/\tilde \nu)} ,
\label{deq}
\end{equation}
where $\tau_d$ is the divergence time and $d_0 = \mu \tau_d$ the expected divergence at neutral sites.

\subsubsection*{Ancestor-directed, outgroup-directed, and corrected frequency spectra}

As explained in the main text, the evolutionary model specified by equations (\ref{upm}) -- (\ref{deq}) serves to relate the  outgroup-directed frequency spectra $q^o (x)$ and the basic frequency distributions of the draft model, $q_\pm (x)$ (equation~\ref{qsel}) without additional fitting parameters. Specifically, the ancestor-directed spectrum $q_s(x; \theta, \nu)$ at synonymous sites, which is of the form~(\ref{qs}), determines the outgroup-directed spectrum
\begin{eqnarray}
q^o_s (x; \theta, \nu) & = & \theta_s \big [ (1 - d_0) Q_0 (x; \nu) + d_0 Q_0 (1-x; \nu) \big ]
\label{qso}
\end{eqnarray}
with the spectral function $Q_0 (x; \nu)$ given by equation~(\ref{Q0}). In other sequence classes, we use ancestor-directed spectra of the form~(\ref{qa}),
\begin{eqnarray}
q (x; \theta, \theta', \zeta, \nu) & = & \theta \, Q_0 (x; \nu) + \theta' \, Q(x; \zeta, \nu)
\nonumber
\\
& = & \theta \, Q_0 (x; \nu) + \theta' \, \big [ \lambda_+ (\zeta, \nu) Q_- (x; \zeta, \nu) + \lambda_- (\zeta, \nu) Q_+ (x; \zeta, \nu) \big ]
\end{eqnarray}
with the spectral functions $Q_\pm (x; \zeta, \nu)$ given by equation~(\ref{Q}) and the allele occupancy probabilities $\lambda_\pm (\zeta, \nu)$ given by equation~(\ref{lambdapm}). These determine the outgroup-directed counterparts
\begin{eqnarray}
q^o(x; \theta, \theta', \zeta, \nu) & = & \theta \big [ (1 - d_0) Q_0 (x; \nu) + d_0 Q_0 (1-x; \nu) \big ]
\nonumber \\
& & +  \theta' \big [ \big (1 - d (\zeta, \nu) \big ) \big (\lambda_+ (\zeta, \nu) Q_- (x; \zeta, \nu) + \lambda_- (\zeta, \nu) Q_+ (x; \zeta, \nu) \big )
\nonumber \\
& &  + d (\zeta, \nu) \big (\lambda_+ (\zeta, \nu) Q_- (1-x; \zeta, \nu) + \lambda_- (\zeta, \nu) Q_+ (1-x; \zeta, \nu) \big ) \big ] .
\label{qao}
\end{eqnarray}

We can reconstruct the synonymous spectrum $q_s(x)$ from $q_s^o (x)$ by inverting the linear map (\ref{qso}),
\begin{equation}
q_s(x) = \frac{1-d_0}{1 - 2d_0} \, q_s^o (x) - \frac{d_0}{1- 2d_0} \, q^o (1-x).
\label{qso_inverse}
\end{equation}
Applying this transformation to the outgroup-polarized spectral data of synonymous mutations, $\hat q_s^o (x)$ (Fig.~\ref{fig_syn}ab), produces the corrected spectral data $\hat q_s(x)$ shown in Supplementary Figure~S2. These provide a {\em bona fide} improved approximation to the underlying spectrum $q_s (x)$. However, the reconstruction becomes noisy in the limit $x \to 1$, where $\hat q_s^o(x)$ is dominated by the component $\hat q_s (1-x)$.

\subsubsection*{Bayesian estimation of model parameters}

Consider a sequence class with population frequency spectrum $q (x; \theta, \theta', \zeta, \nu)$ given by a two-component model of the form~(\ref{qa}); the associated outgroup-polarized spectrum $q^o (x; \theta, \theta', \zeta, \nu)$ is given by equation~(\ref{qao}). In that class, a sample of $n$ random individuals contains mutations of discrete outgroup-polarized frequency $x = k/n$ with probability (see also ref. \cite{Sawyer1992-rp})
\begin{eqnarray}
\tilde q^o (k/n; \theta, \theta', \zeta, \nu) & = &
\left ( \!\!\begin{array}{c} n \\ k \end{array} \!\!\right ) \int_0^1 x^k (1 - x)^{n-k} q^o (x; \theta, \theta', \zeta, \nu) \, dx
\qquad (k = 0, 1, \dots, n);
\label{sampling}
\end{eqnarray}
this expression yields closed analytical expressions involving hypergeometric and Gamma functions.

By calibrating the model distributions $\tilde q (k/n; \theta, \theta', \zeta, \nu)$ with observed site frequency spectra $\hat q^o (k/n)$ and divergence data, we can infer parameters of the model (\ref{qso}) for synonymous sites and of the mixed model (\ref{qao}) for other sequence classes.  Our inference is based on total log likelihood score of the observed frequency counts in a given sequence class,
\begin{equation}
S (\theta, \theta', \zeta, \nu) = L \sum_{k=0}^n \hat q^o (k/n) \log \tilde q^o (k/n; \theta, \theta', \zeta, \nu),
\end{equation}
where $L$ is the total number of sequence sites in the class. We have developed a consistent Bayesian inference scheme that takes into account the allele occupancy (\ref{lambdapm}), the evolutionary dynamics (\ref{deq}), and the sampling statistics (\ref{sampling}). This scheme proceeds in a hierarchical way: we first determine a posterior distribution of parameters $(\theta_s, \nu)$ for synonymous sites, using the single-component model~(\ref{qso}). Then we obtain the posterior distribution of parameters $(\theta_a, \theta'_a, \zeta_a, \nu)$ for amino-acid changes and the analogous distributions for other sequence classes, using the mixed model~(\ref{qa}) with the same value of $\nu$ as for synonymous sites (this constraint does not induce a significant drop in likelihood score).
Our inference scheme is implemented in a software called ``hfit'' \path{https://github.com/stschiff/hfit} using special functions and numerical optimization routines from the Gnu Scientific Library \path{http://www.gnu.org/software/gsl/}, and a custom MCMC algorithm to obtain Maximum Likelihood estimates and confidence intervals for all parameters.

The Bayesian inference scheme, together with the substitution model given by equations (\ref{upm}) -- (\ref{deq}), allows a direct estimate of the rates $u_d, v_b$ and of the interference density $\omega$ from observed frequency spectra and substitutions at synonymous and non-synonymous sites. First, the rate of deleterious mutations in a given sequence class is simply $u_d = \mu \lambda_+ \theta / \theta_s$. Equation~(\ref{qao}) then determines the total rate of deleterious nonsynonymous mutations,
\begin{equation}
u_d = \mu \, \alpha_d =
\mu \, \left (\frac{\lambda_+ (\zeta_a, \nu) \, \theta'_a}{\theta_s} + \frac{\theta_s - \theta_a - \theta'_a}{\theta_s} \right ),
\label{ud}
\end{equation}
which is the sum of contributions from moderately deleterious changes and from strongly deleterious changes. Second, the rate of adaptive amino acid substitutions is given by the excess of nonsynonymous divergence compared to the expectation from the equilibrium model (\ref{deq}),
\begin{equation}
v_b = \mu \, \frac{d_a}{d_s} \, \alpha_b =
\mu \, \frac{d_a}{d_s} \left (1 - d_s \frac{\theta_a}{\theta_s} - d(\zeta_a, \nu) \frac{\theta'_a}{\theta_s} \right ).
\label{vb}
\end{equation}
Here we have treated synonymous mutations as (approximately) neutral. In the local interference regime, we have $\zeta_a \gg   1$ and, hence, $\lambda_+ (\zeta_a, \nu) \approx 1$ and $ d(\zeta_a, \nu) \approx 0$. Equations (\ref{ud}) and (\ref{vb}) then reduce to the expressions given in the main text, $u_d / \mu = \alpha_d$ with $\alpha_d = 1 - (\theta_a /\theta_s)$ and $v_b / \mu = \alpha_b \,(d_a/d_s)$ with $\alpha_b = 1 - (d_s/d_a)(\theta_a/\theta_s)$. These expressions are evaluated using measured divergence data $d_s, d_a$ and maximum-likelihood spectral parameters $\theta_s, \theta_a$. They enter equation (\ref{omega2}) for the interference density $\omega$, which serves to estimate the threshold recombination rate $\rho^*$ from the condition $\omega^* = 1$. To estimate the fraction of adaptive substitutions, $\alpha_b$, in the condensate regime (Fig.~\ref{fig_omega}d), we use the full expression (\ref{vb}).

\subsubsection*{Genomic data and sequence annotation}
We downloaded the complete genome sequences of 168 lines from the \emph{Drosophila Melanogaster Reference Panel (DGRP)} from the DGRP website \path{http://dgrp.gnets.ncsu.edu} and of 27 lines sampled from Rwanda from the \emph{Drosophila Population Genomics Project} \path{http://dpgp.org} as fasta files. We downloaded the reference sequences from Drosophila simulans and from \emph{Drosophila yakuba}, aligned to the reference sequence of \emph{Drosophila melanogaster} from the UCSC genome browser (\path{https://genome.ucsc.edu}). For both outgroups, we compute outgroup-directed allele frequencies at all sites at which (i) there is a valid outgroup allele, and (ii) at least 150 lines of the DGRP sequences or 25 lines of the DPGP sequences have a called allele. We then downsample all sites to 25 called alleles, using random sampling without replacement (hypergeometric sampling).

We downloaded gene annotations from flybase [43]. We define annotation categories as follows. {\bf Intergenic}: intergenic regions that are at least 5kb away from genes, {\bf Intron}: introns of protein-coding genes, {\bf UTR}:  untranslated regions in exons, {\bf Synonymous}: protein-coding sites of the reference genome at which none of the three possible point mutation changes the encoded amino acid, {\bf Nonsynonymous}: protein-coding sites on the reference at which any of the three possible point mutations changes the encoded amino acid. Most genes have multiple associated transcripts due to alternative splicing. We choose the transcript corresponding to the longest encoded protein coding sequence for each gene and annotated introns, UTRs, synonymous and nonsynonymous sites according to that transcript. See Supplementary Table S1 for the number of sites in a given annotation category on the different chromosomes.

Maps of mean recombination rates within 100kb windows were obtained from Comeron et al. \cite{Comeron2012-jh} through the website \path{http://www.recombinome.com}. We use the recombination map to annotate every site in the Drosophila genome. We then use only synonymous sites on the autosomes (2L, 2R, 3L and 3R) and define quantile boundaries on this set. Specifically, we sort all recombination rate values of this set of sites and determine recombination rate bins by dividing the data set into 21 equally large subsets of values. We then use these quantile boundaries to bin all sites (not just synonymous sites) into bins according to their local recombination rate. The quantile boundaries used in this study for autosomal data are (in cM/Mb):  0.0, 0.069, 0.217, 0.415, 0.44, 0.821, 1.055, 1.29, 1.415, 1.592, 1.741, 1.938, 2.169, 2.354, 2.612, 2.838, 3.156, 3.461, 3.796, 4.244, 5.395, Infinity. All binned allele frequency data is given in Supplementary Table S1.

\subsubsection*{Simulations of evolutionary processes in recombining populations}

We use the SLiM simulator \cite{Messer2013-wc} to simulate a population of sequences evolving under mutations, drift, selection, and recombination; the genome of each individual has $100,000$ sites. To mimic the {\em Drosophila} phylogeny, we start from a single population of size $N=1000$ that evolves for $10,000$ generations, then splits into ingroup and outgroup populations of size $N=1000$; these evolve in isolation for another $10,000$ generations. Finally, we sample one individual from the outgroup population and 20 individuals from the ingroup population.

We consider three classes of mutations: neutral mutations, beneficial mutations and deleterious mutations, the latter two with fixed selection coefficient $s_\mathrm{ad}=0.01$. The rate of neutral mutations is $\mu = 1.5\times 10^{-6}$, the rate of beneficial mutations varies from $u_b = 0$ to $2.5\times 10^{-7}$, and the rate of deleterious mutations from $u_d = 0$ to $3\times 10^{-6}$. We also run simulations with only one class of selected mutations (i.e., $u_d = 0$ or $u_b = 0$). The recombination rate $\rho$ varies in the range $ 10^{-7}$ to $10^{-4}$.

We use these simulations to display the transition from local interference to the interference condensate and to corroborate our scaling theory (Fig.~\ref{fig_sim}). In particular, the simulations demonstrate that the maximum-likelihood shape parameter $\nu^*$ inferred from the spectral data of synonymous sequence sites can serve as a faithful marker of the interference condensate regime.

\section*{Acknowledgments}
We would like to thank P.W. Messer for comments on an earlier version of the manuscript.

\section*{Legends for Supplementary Tables}

\textbf{Supplementary Table~S1: Drosophila site frequency spectra.} The table lists the site frequency spectrum data that we analyse here. The data is separated into five annotation classes and 21 recombination bins.

\bigskip \noindent
\textbf{Supplementary Table~S2: Parameter estimates.} The table lists all parameter estimates obtained from Drosophila data using our draft model.

\renewcommand\thefigure{S\arabic{figure}}
\setcounter{figure}{0}

\begin{figure}
  \includegraphics[width=\textwidth]{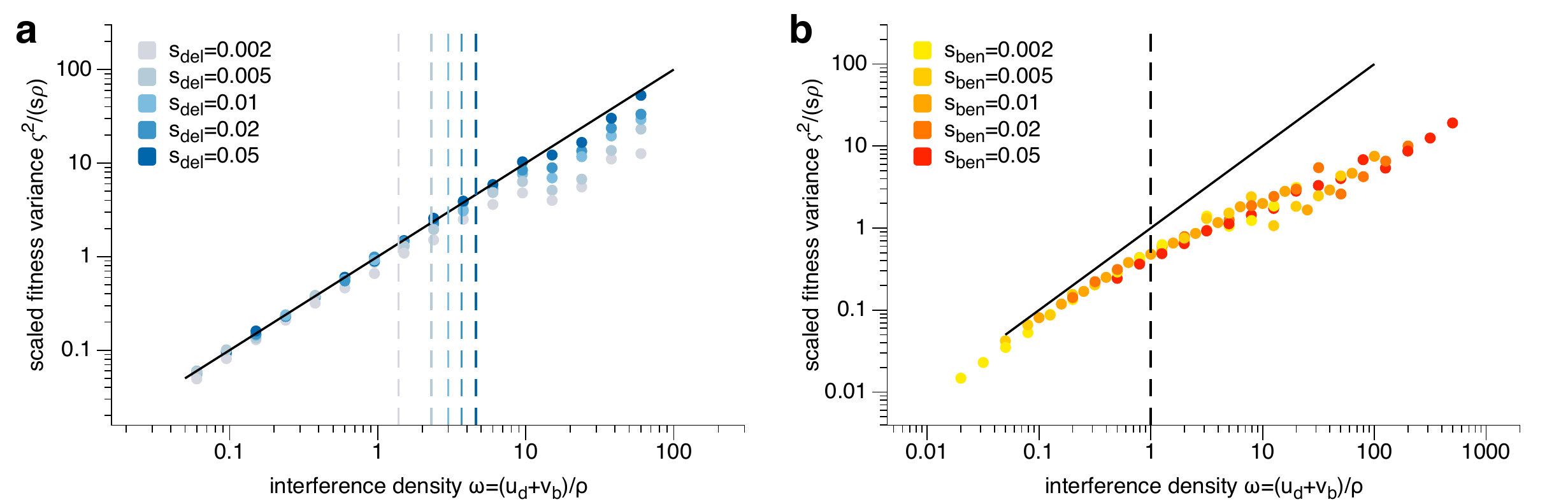}
  \caption{\textbf{Extensions of scaling theory.} The scaled fitness variance per site, $\varsigma^2 / (s \rho)$, is plotted against the interference density $\omega$ for simulated evolution with two complementary distributions of selective effects. (a) Background selection. Evolution under solely deleterious mutations with rate $u_d = 3\times 10^{-6}$ and a single-valued selection coefficient $(-s)$ (marked by color); other simulation parameters as in Fig.~\ref{fig_sim}. The onset point of interference selection, $\omega^*$, increases with $s$, in qualitative agreement with equation~(\ref{match}) (dashed lines indicate the predicted value of $\omega^*$). (b) Selective sweeps. Evolution under solely beneficial mutations with rate $u_d = 2.5 \times 10^{-7}$ and a single-valued selection coefficient $s$ (marked by color); other simulation parameters as in Fig.~\ref{fig_sim}. The onset point of interference selection, $\omega^* \sim 1$, is independent of $s$, as predicted by equations~(\ref{omega}) and~(\ref{varsigma2}). Both selection scenarios have been studied in previous work \cite{Hudson1995-kx,Nordborg1996-xc,Weissman2012-of} but are not realistic assumptions for {\em Drosophila}.}
  \label{fig_S_extensions}
\end{figure}

\begin{figure}
  \includegraphics[width=\textwidth]{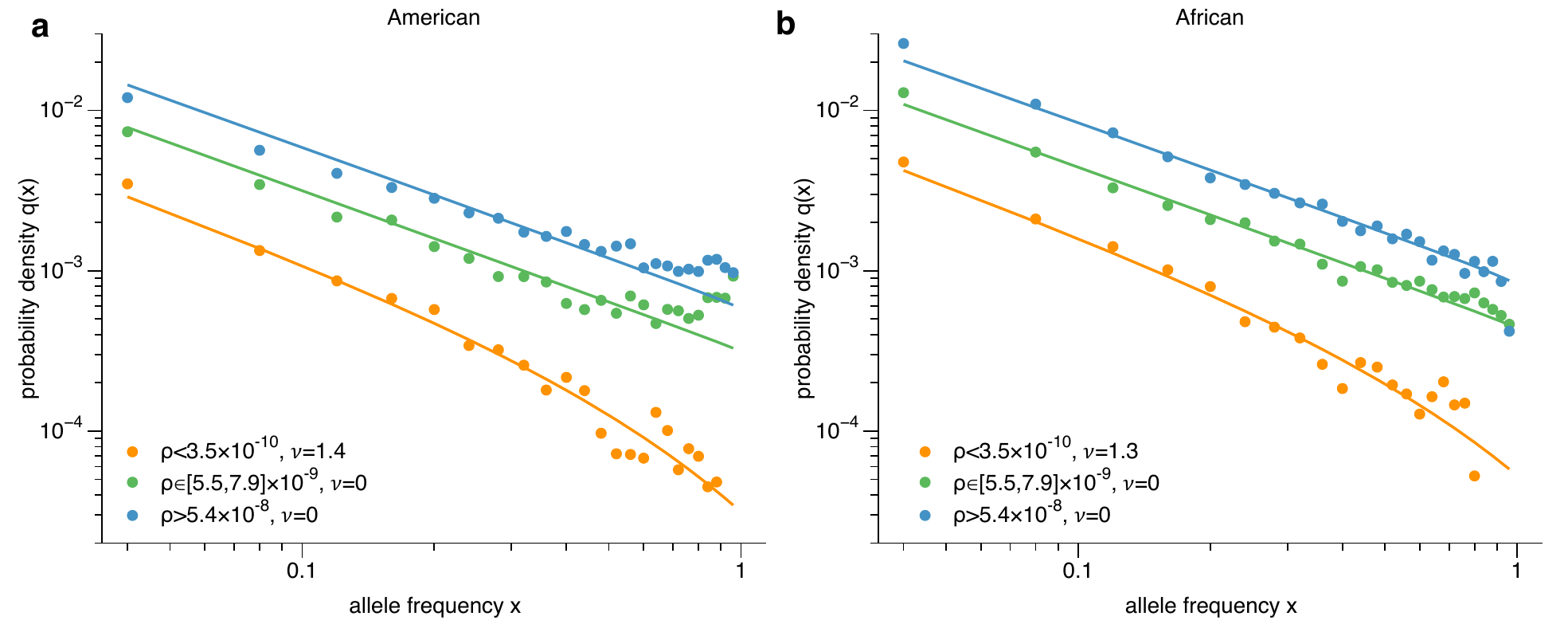}
  \caption{\textbf{Corrected frequency spectra of synonymous mutations.} Ancestor-directed empirical spectra $\hat q_s (x)$ for two populations of {\em D. melanogaster} are reconstructed from the outgroup-directed spectra $\hat q_s^o (x)$ (Fig.~\ref{fig_syn}ab) by the inverse linear map~(\ref{qso_inverse}).}
  \label{fig_S_correctedSpectra}
\end{figure}

\begin{figure}
  \includegraphics[width=\textwidth]{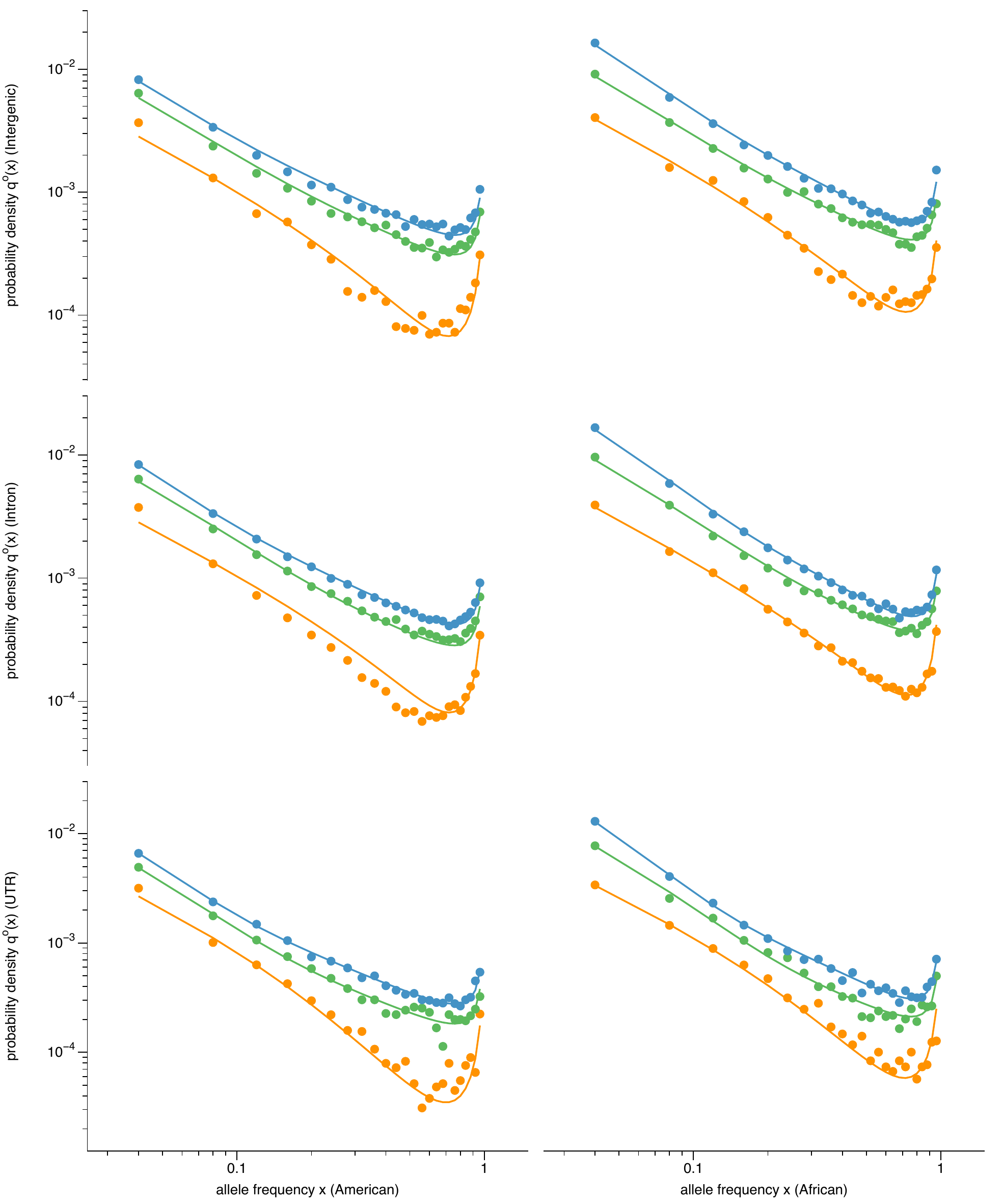}
  \caption{\textbf{Frequency spectra of mutations in UTR, introns and intergenic sequence.} Outgroup-polarized sample spectra $\hat q^o(x)$  from two populations of \emph{D. melanogaster}, together with maximum-likelihood spectra $q^o(x)$. We use the shape parameter $\nu$ inferred from synonymous mutations (Fig.~\ref{fig_syn}c) and a two-component model of the same form as for nonsynonymous mutations; see equations (\ref{qa}) and~(\ref{qao}).}
  \label{fig_S_otherSpectra}
\end{figure}

\bibliography{bibliography}{}

\end{document}